\RequirePackage{fix-cm}
\documentclass[english,aps,prb,twocolumn,showpacs,superscriptaddress,groupaddress,floatfix,graphics,graphicx]{revtex4-2}
\usepackage[LGR,T1]{fontenc}
\usepackage[latin9]{inputenc}
\usepackage{xcolor}
\usepackage{babel}
\usepackage{float}
\usepackage{units}
\usepackage{amsmath}
\usepackage{amssymb}
\usepackage{graphicx}
\usepackage[unicode=true,pdfusetitle,
 bookmarks=true,bookmarksnumbered=false,bookmarksopen=false,
 breaklinks=true,pdfborder={0 0 0},pdfborderstyle={},backref=false,colorlinks=true]
 {hyperref}
\hypersetup{
 colorlinks,linkcolor=blue,citecolor=blue,urlcolor=blue}

\makeatletter

\DeclareRobustCommand{\greektext}{%
  \fontencoding{LGR}\selectfont\def\encodingdefault{LGR}}
\DeclareRobustCommand{\textgreek}[1]{\leavevmode{\greektext #1}}
\ProvideTextCommand{\~}{LGR}[1]{\char126#1}

\makeatother

\begin{document}
\global\long\def\vect#1{\overrightarrow{\mathbf{#1}}}%

\global\long\def\abs#1{\left|#1\right|}%

\global\long\def\av#1{\left\langle #1\right\rangle }%

\global\long\def\ket#1{\left|#1\right\rangle }%

\global\long\def\bra#1{\left\langle #1\right|}%

\global\long\def\tensorproduct{\otimes}%

\global\long\def\braket#1#2{\left\langle #1\mid#2\right\rangle }%

\global\long\def\omv{\overrightarrow{\Omega}}%

\global\long\def\inf{\infty}%

\title{Nodal vacancy bound states and resona\textcolor{black}{nces in three-dimensional
W}eyl semimetals}
\author{J. P. Santos Pires}
\email{up201201453@fc.up.pt}

\address{Departamento de Física e Astronomia, Faculdade de Ciências da Universidade
do Porto, Rua do Campo Alegre, s/n, 4169-007 Porto, Portugal}
\address{Centro de Física das Universidades do Minho e do Porto (CF-UM-UP)
and Laboratory of Physics for Materials and Emergent Technologies
LaPMET, University of Porto, 4169-007 Porto, Portugal}
\author{S. M. João}
\address{Departamento de Física e Astronomia, Faculdade de Ciências da Universidade
do Porto, Rua do Campo Alegre, s/n, 4169-007 Porto, Portugal}
\address{Centro de Física das Universidades do Minho e do Porto (CF-UM-UP)
and Laboratory of Physics for Materials and Emergent Technologies
LaPMET, University of Porto, 4169-007 Porto, Portugal}
\author{Aires Ferreira}
\address{Department of Physics and York Centre for Quantum Technologies, University
of York, YO10 5DD, York, United Kingdom}
\author{B. Amorim}
\address{Centro de Física das Universidades do Minho e do Porto (CF-UM-UP)
and Laboratory of Physics for Materials and Emergent Technologies
LaPMET, Universidade do Minho, 4710-057 Braga, Portugal}
\author{J. M. Viana Parente Lopes}
\email{jlopes@fc.up.pt}

\address{Departamento de Física e Astronomia, Faculdade de Ciências da Universidade
do Porto, Rua do Campo Alegre, s/n, 4169-007 Porto, Portugal}
\address{Centro de Física das Universidades do Minho e do Porto (CF-UM-UP)
and Laboratory of Physics for Materials and Emergent Technologies
LaPMET, University of Porto, 4169-007 Porto, Portugal}
\begin{abstract}
The\textcolor{black}{{} electronic structure o}f a cubic $\mathcal{T}$-symmetric
Weyl semimetal is analyzed in the presence of atomic-sized vacancy
defects. Isolated vacancies are shown to generate nodal bound states
with $r^{{\scriptscriptstyle -2}}$ asymptotic tails, even when immersed
in a weakly disordered environment. These states show up as a significantly
enhanced nodal density of states which, as the concentration of defects
is increased, reshapes into a nodal peak that is broadened by inter-vacancy
hybridization into a comb of satellite resonances at finite energies.
Our results establish point defects as a crucial source of elastic
scattering that leads to nontrivial modifications in the electronic
structure of Weyl semimetals.
\end{abstract}
\maketitle

\section{Introduction}

\noindent With the advent of three-dimensional (3D) topological insulators\,\citep{Fu2007,Hasan2010},
the search for topological semimetals emerging at the transition between
gapped phases of matter has flourished. Rather than being a fine-tuned
situation, it was envisaged by Murakami\,\citep{murakami2007} that,
without inversion symmetry, topological phase transitions can proceed
through an intermediate stage, in which a pair of two-fold degenerate
band-crossing points moves around the first Brillouin zone until it
finally merges together and gives rise to a new gapped phase. Such
a stable gapless state was dubbed a Weyl semimetal (WSM)\,\citep{Halasz2012,Armitage2018}
because low-energy excitations around these band-crossings are described
by a decoupled pair of ($3\!+\!1)$--dimensional Weyl equations of
opposite chirality\,\citep{Weyl1929}. Later on, such a topological
gapless phase was also shown to be possible in centrossymetric crystals,
so long as time-reversal symmetry is broken\,\citep{Wan2011,Xu2011,Burkov2011,Koshino2016}
(magnetic WSM). Crucially, in all cases, the band-crossings form pairs
of point-like sources (or sinks) of Berry flux in momentum space,
analogous to the well-known ``diabolical points'' described by Berry\,\citep{Berry1984}
in a generic two-level quantum system. Therefore, isolated Weyl nodes
are topologically protected degeneracies in the electronic band structure
that are robust to parametric changes of the Hamiltonian.

The topological character of WSMs yields important physical consequences,
from the existence of surface Fermi arcs\,\citep{Wan2011,Hosur2012,Witten2015,Haldane2014,Hashimoto2017}
that connect Weyl nodes in the surface-projected first Brillouin zone
(fBZ), to the remarkable condensed matter realization of QED's chiral
anomaly\,\citep{Adler1969,Bell1969}. The latter drives distinctive
unconventional transport effects, such as a negative longitudinal
magnetoresistance\,\citep{Son2013,Zhang2016}, a giant in-plane Hall
effect\,\citep{Burkov2017,Nandy2017} and the chiral magnetic effect\,\citep{Nielsen83}.
Nonetheless, perhaps the most remarkable property of a WSM is its
resilience to the effects of unavoidable perturbations, such as disorder
or crystal defects. From a theoretical standpoint, the study of disorder
effects in both spectral\,\citep{Bera2016,Pixley2015,Pixley2016,Wilson2020,Pires2021,Pixley2021}
and transport properties\,\citep{Fradkin1986,Roy2014,Syzranov2015,Slager17,Shen2020}
of WSMs have been the subject of intense research. A big focus was
placed on the effects of random potentials that can yield non-Anderson
quantum criticality at a finite disorder strength\,\citep{Fradkin1986,Syzranov2018,Roy18}.
As the system is driven through this critical point, the semi-metallic
character of the nodal single-electron states gets destroyed long
before they become exponentially localized at the Anderson transition\,\citep{Pixley2015}.

\vspace{-0.08cm}

In contrast to conventional Anderson transitions \citep{Janssen1994,Evers2008},
the disorder-averaged nodal density of states (nDoS) in disordered
WSMs is deemed an appropriate order parameter by field-theoretical
calculations\,\citep{Shindou2009,Goswami2011,Roy2014,Syzranov2015_2},
as well as the numerical observation of its sharp power-law growth
above some critical disorder strength\,\citep{Kobayashi2014,Pixley2015,Pixley2016}.
However, recent studies of nonperturbative instantonic effects have
revealed that rare disorder configurations lift the nDoS and round-out
its critical behavior \citep{Nandkishore2014,Gurarie2017,Pixley2021},
thus challenging the conventional scenario. A physical picture was
then put forward by Nandkishore \textit{et al}\,\citep{Nandkishore2014},
who associated the nDoS lift to smooth rare-regions of a random potential
landscape that can sporadically bound eigenstates at the nodal energy,
giving way to an exponentially small but \emph{nonzero} nDoS. Despite
being a controversial proposal\,\citep{Pixley2016,Buchhold2018,Buchhold2018_2,Ziegler2018,Wilson2020,Pires2021},
the \emph{avoided} quantum criticality due to rare events was eventually
confirmed in subsequent numerical studies\,\citep{Pixley2016,Wilson2020,Pires2021}
and it is now believed to be the most general scenario\,\citep{Pixley2021}.

\textcolor{black}{In spite of these numerous theoretical studies regarding
the effects of random potential disorder, very little is known about
the role played by point defects and other common disorder sources.
Currently, time-reversal symmetric Weyl fermions can be realized as
low-energy quasiparticles in a myriad of materials, most notably within
the TaAs cubic family (also including NbAs, TaP and NbP)\,\citep{Lv21},
which can be grown as single crystals using chemical vapor transport
techniques\,\citep{Ghimire2015}. In the growth process, lattice
defects are likely to form\,\citep{Buckeridge2016,Gen2016} and,
as demonstrated in previous experimental studies based on transmission
electron microscopy\,\citep{Besara2016} and Raman scattering\,\citep{Liu2016},
even high-quality samples generally host a considerable density of
defects, mostly vacancies and stacking faults. Adding to their natural
occurrence, vacancy defects can also be artificially induced by means
of particle irradiation\,\citep{Zhang2019,Fu2020}, a well-tested
technique previously used to generate defects in graphene\,\citep{Lehtinen2010}
and two-dimensional semiconductors\,\citep{Lin2016,Bertoldo2022}.
Since point defects can significantly change the electronic structure
of materials, a study of their impact as a source of disorder in WSMs
opens up interesting possibilities. Promising results were reported
by Xing }\textit{\textcolor{black}{et al.}}\textcolor{black}{\,\citep{Xing2020},
where atomic vacancies hosted by the (magnetic) WSM $\text{Co}_{3}\text{Sn}_{2}\text{S}_{2}$
were linked to the presence of exotic localized spin--orbit polaron
states on its surface. In this paper, we push this line forward by
theoretically analyzing the electronic properties of Weyl fermions
in the presence of point defects. More specifically, we characterize
the electronic wavefunctions and corresponding density of states (DoS)
of a lattice $\mathcal{T}$-symmetric Weyl semimetal with finite concentrations
of randomly distributed atomic-sized vacancies with one (}\textit{\textcolor{black}{half-vacancy}}\textcolor{black}{)
or two (}\textit{\textcolor{black}{full-vacancy}}\textcolor{black}{)
orbitals missing from the defect sites.}

The remainder of this paper is organized as follows. In Sec.\,\ref{sec:Single Vacancy Lattice},
we introduce our working model and a \textit{projected Green's function
formalism} (pGF) that is used to calculate the vacancy-induced DoS
deformation and show that algebraically decaying nodal bound states
appear for isolated half- and full-vacancies. In Sec.\,\ref{sec:Robustness to Disorder},
the existence of nodal bound states is further verified by \textit{Lanczos
diagonalization}\,\citep{Lanczos50,paige1980,Lehoucq97} (LD) of
lattices containing an isolated vacancy. The robustness of these states
to an additional weakly disordered environment is also discussed.
In Sec.\,\ref{sec:Finite-Vacancy-Concentrations}, we analyze the
averaged DoS of a WSM with a finite concentration of vacancies, employing
a combination of LD and \textit{spectral methods} \citep{Weise2006,Ferreira2015,Joao2020}.
While confirming that localized eigenstates still appear and enhance
the value of the DoS around the Weyl node, our results further show
that inter-vacancy hybridization quickly broadens the nodal peak in
the DoS, forming a comb of symmetrically-placed subsidiary sharp resonances
for a moderate concentration of defects. Finally, Sec.\,\ref{sec:Conclusion-and-Outlook}
summarizes our key results and gives an outlook.

\vspace{-0.6cm}

\section{\label{sec:Single Vacancy Lattice}Modeling an Isolated Vacancy in
a Weyl Semimetal}

\vspace{-0.2cm}

\begin{figure}[t]
\vspace{-0.2cm}
\begin{centering}
\hspace{-0.2cm}\includegraphics[scale=0.27]{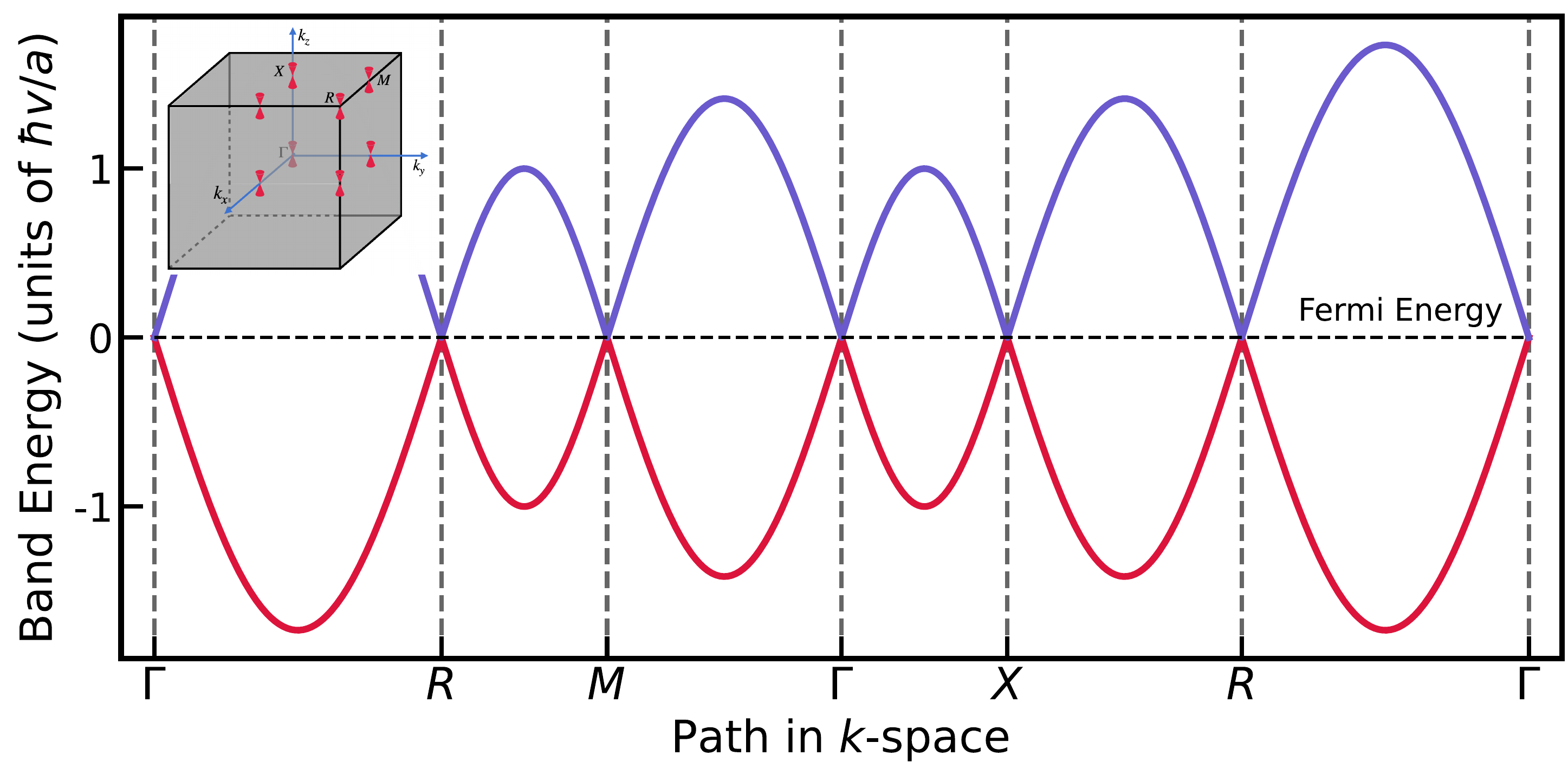}
\par\end{centering}
\vspace{-0.3cm}

\caption{\label{fig:Squeme-of-the}Band structure of the clean\textcolor{black}{{}
lattice WSM model along the $\boldsymbol{k}$-space path indicated
in the inset. The locations of the eight no}n-equivalent Weyl cones
are represented as well.}

\vspace{-0.6cm}
\end{figure}

A lattice vacancy is a common crystalline defect\,\citep{Siegel78}.
When a crystal is formed some sites are not properly occupied by the
corresponding atoms, creating a proportion of vacant sites\,\citep{Crawford72,Schaefer99}
that act as a source of disorder. In the language of tight-binding
Hamiltonians, a vacancy can be modeled by removing one or more Wannier
orbitals from a randomly chosen lattice site. We start by determining
the effects of introducing a single lattice \textit{half-} or\textit{
full-vacancy} in a two-band model of a WSM. We employ a particle-hole
symmetric model that lives in a simple cubic lattice ($\mathcal{L}$)
and features a low-energy dispersion relation with eight isotropic
Weyl nodes pinned to the time-reversal invariant momenta of the cubic
fBZ (see Fig.\,\ref{fig:Squeme-of-the}). The lattice Hamiltonian\,\citep{Pixley2021}
may be written as 

\vspace{-0.6cm}
\begin{equation}
\mathcal{H}_{0}\!=\!\frac{\hbar v}{2a}\sum_{\mathbf{R}\in\mathcal{L}}\sum_{{\scriptscriptstyle i=x\!,\!y\!,\!z}}\!\!\left[\Psi_{\mathbf{R}}^{\dagger}\!\cdot\!\sigma^{i}\!\!\cdot\!\Psi_{\mathbf{R}+a\hat{\mathbf{x}}_{i}}\!-\!\text{h.c.}\right]\label{eq:HamiltonianLattice}
\end{equation}
where $a$ is the lattice parameter, $v$ is the Fermi velocity, $\hat{\mathbf{x}}_{i}\!=\!(\hat{x},\hat{y},\hat{z})$
are cartesian unit vectors, $\boldsymbol{\sigma}$ is the vector of
$2\!\times\!2$ Pauli matrices and $\Psi_{\mathbf{R}}^{\dagger}\!=\![c_{\mathbf{R},1}^{\dagger},c_{\mathbf{R},2}^{\dagger}]$
is a local two-orbital fermionic creation operator. \textcolor{black}{Equipped
with this lattice description, the vacancy defects are implemented
in two distinct ways. In our pGF calculations below, lattice vacancies
are created by canceling all hoppings at the defect site, which leaves
behind uncoupled zero energy Wannier states. In contrast, when the
system is analyzed using spectral methods or LD {[}Secs.\,\ref{sec:Robustness to Disorder}
and \ref{sec:Finite-Vacancy-Concentrations}{]}, the Hilbert space's
dimension is effectively reduced by iterating with vectors orthogonal
to the removed orbitals.}

\vspace{-0.5cm}

\subsection{Clean Lattice Green Function and Nodal Point Symmetries}

\vspace{-0.3cm}

Before diving into the analysis of the electronic structure of WSMs
with vacancy defects, we first establish some basic results. The Bloch
Hamiltonian of the clean system admits the simple representation 

\vspace{-0.3cm}

\begin{equation}
\mathcal{H}_{0}(\mathbf{k})\!=\!i\hbar v\boldsymbol{\sigma}\!\cdot\!\boldsymbol{\sin}\,a\mathbf{k},
\end{equation}
with $\boldsymbol{\sin}\,a\mathbf{k}\equiv\left(\sin ak_{x},\sin ak_{y},\sin ak_{z}\right)$,
and which yields the dispersion relation represented in Fig.\,\ref{fig:Squeme-of-the}.
The clean retarded lattice Green's function (LGF), defined formally
as $G_{0}(E,\mathbf{R}_{j}\!-\!\mathbf{R}_{i})\!=\!\left[E\!+\!i0^{{\scriptscriptstyle +}}\!\!\!-\!\mathcal{H}_{0}\right]_{\mathbf{R}_{j},\mathbf{R}_{i}}^{{\scriptscriptstyle -1}}$,
can be written, in terms of dimensionless quantities, as 

\vspace{-0.5cm}
\begin{equation}
\boldsymbol{G}_{0}(\varepsilon,\boldsymbol{\Delta R})\!=\!\!\!\int_{{\scriptscriptstyle \![-\!\pi,\pi]^{3}}}\!\!\!\frac{d^{{\scriptscriptstyle \text{(\!3\!)}}}\boldsymbol{k}}{8\pi^{3}}\frac{\varepsilon\!+\!i\boldsymbol{\sigma}\!\cdot\!\boldsymbol{\sin k}}{\varepsilon^{2}\!+\!\abs{\boldsymbol{\sin k}}^{2}}e^{-i\mathbf{k}\cdot\boldsymbol{\Delta R}},\label{eq:GreenFunctionClean}
\end{equation}
where $\varepsilon\!=\!Ea/\hbar v\!+\!i\eta$ is the dimensionless
energy (shifted by an imaginary amount $\eta$), $\boldsymbol{k}$
is the crystal momentum (in units of $a^{{\scriptscriptstyle -1}}$),
$\boldsymbol{\Delta R}\!=\!\left(n_{x},n_{y},n_{z}\right)$, and $n_{i}\!\in\!\mathbb{Z}$
are indices that label particular sites. Equation\,(\ref{eq:GreenFunctionClean})
can be expressed in terms of four basic integrals over the domain
$[-\pi,\pi]^{3}$, \textit{i.e.},

\vspace{-0.4cm}

\textcolor{black}{
\begin{equation}
\boldsymbol{G}_{0}(\varepsilon,\!\boldsymbol{\Delta R})\!=\!\mathcal{I}_{\varepsilon}^{0}\!\!\left(\boldsymbol{\Delta R}\right)-\!\!\!\!\!\sum_{j=x,y,z}\!\!\!\!i\sigma^{j}\,\mathcal{I}_{\varepsilon}^{j}\!\left(\boldsymbol{\Delta R}\right),\label{eq:G0Structure}
\end{equation}
}where the complex-valued integrals are 
\begin{figure}[t]
\vspace{-0.2cm}
\begin{centering}
\includegraphics[scale=0.16]{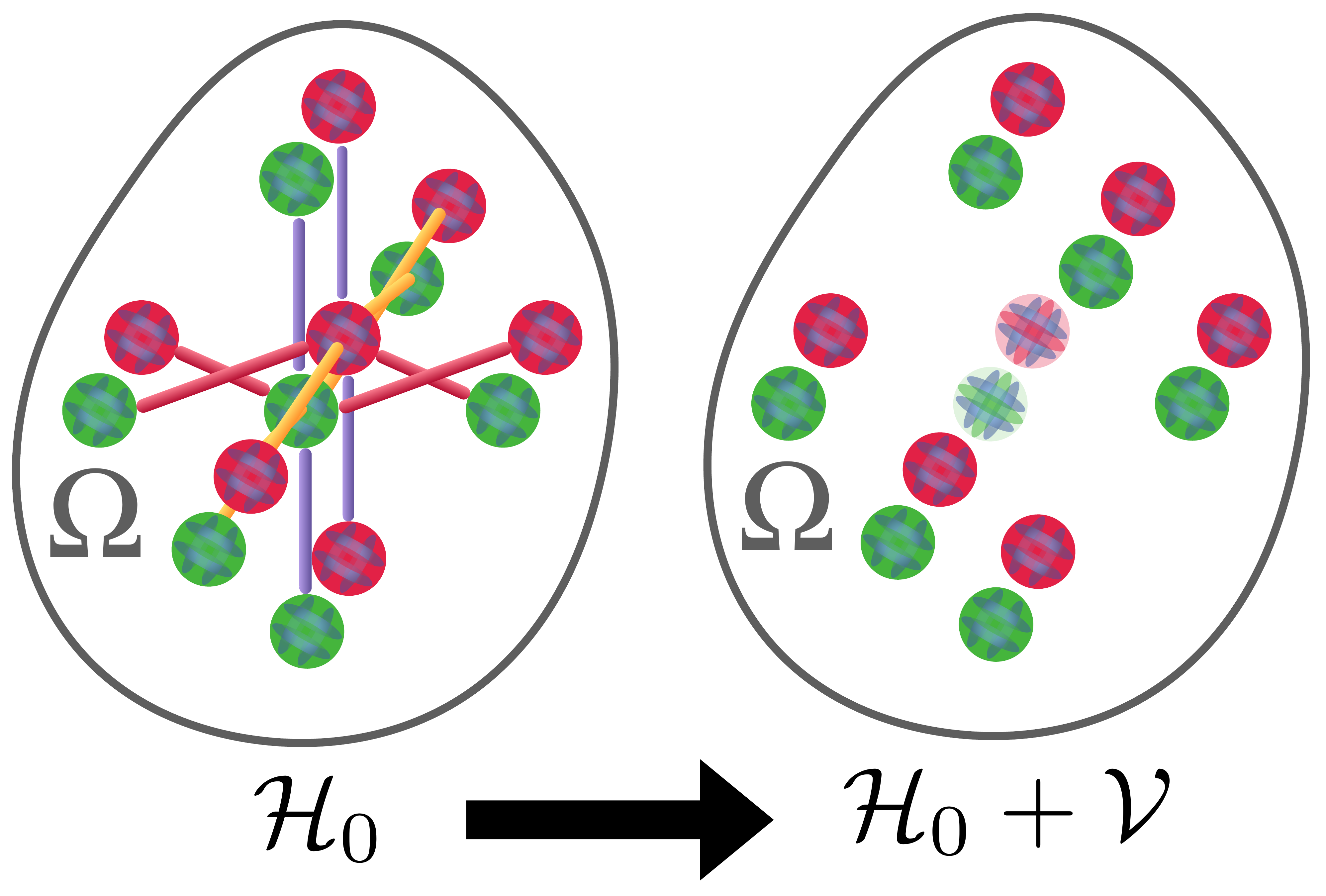}
\par\end{centering}
\vspace{-0.3cm}

\caption{\label{fig:Cluster}Scheme of the local perturbation defined in Eq.\,(\ref{eq:Perturbation}).}

\vspace{-0.4cm}
\end{figure}

\vspace{-0.3cm}

\begin{subequations}
\begin{align}
\mathcal{I}_{\varepsilon}^{0}\!\left(\!\boldsymbol{\Delta R}\right)\! & =\!\!\int_{{\scriptscriptstyle \![-\!\pi,\pi]^{3}}}\!\!\!\!d^{{\scriptscriptstyle \text{(\!3\!)}}}\!\boldsymbol{k}\frac{\varepsilon\,e^{ik_{x}n_{x}}e^{ik_{y}n_{y}}e^{ik_{z}n_{z}}}{8\pi^{3}\left(\varepsilon^{2}+\abs{\boldsymbol{\sin k}}^{2}\right)}\,,\label{eq:Integrals}\\
\mathcal{I}_{\varepsilon}^{x}\!\left(\!\boldsymbol{\Delta R}\right)\! & =\!\!\!\int_{{\scriptscriptstyle \![-\!\pi,\pi]^{3}}}\!\!\!\!\!d^{{\scriptscriptstyle \text{(\!3\!)}}}\!\boldsymbol{k}\frac{\sin\!k_{x}e^{ik_{x}n_{x}}e^{ik_{y}n_{y}}e^{ik_{z}n_{z}}}{8\pi^{3}\left(\varepsilon^{2}+\abs{\boldsymbol{\sin k}}^{2}\right)},\label{eq:Integrals2}
\end{align}
\end{subequations}

\noindent with $\mathcal{I}_{\varepsilon}^{y,z}\!\left(\!\boldsymbol{\Delta R}\right)$
being obtained from $\mathcal{I}_{\varepsilon}^{x}\!\left(\!\boldsymbol{\Delta R}\right)$
by a cyclic permutation of the set $(n_{x},n_{y},n_{z})$. These integrals
can be numerically evaluated with high precision (see Appendix\,\ref{sec:The-Clean-Lattice})
and display useful symmetry properties. First, the parity of the integrand
dictates that $\boldsymbol{G}_{0}(E,\boldsymbol{\Delta R})\!=\!0$
whenever the set $(n_{x},n_{y},n_{z})$ contains more than one odd
integer. This small-scale property of the LGF can be traced back to
the existence of eight non-equivalent valleys which are specific to
this lattice model. Additionally, there are non-spatial symmetries
which can be deduced from Eqs.\,(\ref{eq:Integrals})-(\ref{eq:Integrals2}),
most notably

\vspace{-0.3cm}

\begin{subequations}
\begin{align}
\mathcal{I}_{\varepsilon}^{0}\!\left(\!\boldsymbol{\Delta R}\right)\! & =\!\left[\mathcal{I}_{\varepsilon^{*}}^{0}\!\left(\!\boldsymbol{\Delta R}\right)\right]^{*}\label{eq: symmetry 1}\\
\mathcal{I}_{\varepsilon}^{j}\!\left(\!\boldsymbol{\Delta R}\right)\! & =\!-\!\left[\mathcal{I}_{\varepsilon^{*}}^{j}\!\left(\!\boldsymbol{\Delta R}\right)\right]^{*}\label{eq: symmetry 2}
\end{align}
\end{subequations}

\noindent and also

\vspace{-0.4cm}

\begin{equation}
\text{Re}\!\left[\mathcal{I}_{i0^{+}}^{0}\!\left(\!\boldsymbol{\Delta R}\right)\right]\!=\!\!\!\!-\lim_{\eta\to0^{+}}\!\int_{\text{[-\ensuremath{\pi},\textgreek{p}]}^{3}}\!\!\!\!\!\!\!\!\!\!\!d^{{\scriptscriptstyle \text{(\!3\!)}}}\boldsymbol{k}\frac{\eta\,\sin\left(\boldsymbol{k}\cdot\boldsymbol{\Delta R}\right)}{8\pi^{3}\left(\eta^{2}\!+\!\abs{\boldsymbol{\sin k}}^{2}\right)}\!=\!0,
\end{equation}
by employing the cubic symmetry of the fBZ. Together, these three
properties imply that the LGF at $E\!=\!0$ can be represented in
the simple form

\vspace{-0.5cm}
\begin{equation}
\boldsymbol{G}_{0}(0,\!\boldsymbol{\Delta R})\!=\!\!\!\!\!\!\sum_{j=x,y,z}\!\!\!\!\sigma^{j}\text{Im}\!\left[\mathcal{I}_{i0^{+}}^{j}\!\!\left(\boldsymbol{\Delta R}\right)\right],\label{eq:G(E=00003D0)}
\end{equation}
which is non-zero if and only if $\boldsymbol{\Delta R}\!=\!(n_{x},n_{y},n_{z})$
features a single odd integer.

\vspace{-0.3cm}

\subsection{Projected Green's Function for a Lattice Vacancy}

\vspace{-0.2cm}

\noindent Within a lattice description, a vacancy can be modeled by
removing hoppings connecting one (or several) orbitals within a unit
cell to its neighbors. With no loss of generality, let us consider
a vacancy at the origin, $\boldsymbol{R}\!=\!\boldsymbol{0}$. The
Hamiltonian is then $\mathcal{H}\!=\!\mathcal{H}_{0}\!+\!\mathcal{V}$
with

\vspace{-0.55cm}\textcolor{black}{
\begin{align}
\mathcal{V}\! & =\!-\frac{\hbar v}{2a}\!\!\sum_{{\scriptscriptstyle i=x\!,\!y\!,\!z}}\!\!\left[\Psi_{\boldsymbol{0}}^{\dagger}\!\cdot\!\sigma^{i}\!\!\cdot\!\Psi_{\boldsymbol{0}+a\hat{\mathbf{x}}_{i}}\!\!-\!\Psi_{\boldsymbol{0}}^{\dagger}\!\cdot\!\sigma^{i}\!\!\cdot\!\Psi_{\boldsymbol{0}-a\hat{\mathbf{x}}_{i}}\!\!-\!\text{h.c.}\right].\label{eq:Perturbation}
\end{align}
}

\vspace{-0.25cm}

\noindent This operator has the advantage of having a finite support\textit{,
i.e.}, it acts only on sites $\Omega\!=\!\left\{ \mathbf{0},\pm a\hat{\mathbf{x}}_{1},\pm a\hat{\mathbf{x}}_{2},\pm a\hat{\mathbf{x}}_{3}\right\} $
that form the octahedron shown in Fig.\,\ref{fig:Cluster}. Such
local perturbations to a lattice model can be conveniently studied
by using the pGF method. Treating $\mathcal{V}$ as a perturbation,
we obtain a set of Dyson's equations for the system's Green's function,
$\boldsymbol{G}(\varepsilon)$, in the presence of the vacanc\textcolor{black}{y,
}\textit{\textcolor{black}{i.e.}}\textcolor{black}{,}

\vspace{-0.5cm}

\begin{subequations}
\begin{align}
\boldsymbol{G}(\varepsilon) & \!=\!\boldsymbol{G}_{0}(\varepsilon)\!+\!\boldsymbol{G}(\varepsilon)\!\cdot\!\mathcal{V}\!\cdot\!\boldsymbol{G}_{0}(\varepsilon)\\
\boldsymbol{G}(\varepsilon) & \!=\!\boldsymbol{G}_{0}(\varepsilon)\!+\!\boldsymbol{G}_{0}(\varepsilon)\!\cdot\!\mathcal{V}\!\cdot\!\boldsymbol{G}(\varepsilon),
\end{align}
\end{subequations}

\noindent where $\cdot$ denotes the matrix product defined in the
full Hilbert space. To solve these equations, we proceed in two steps:
\textit{i)} By projecting them into $\Omega$, we can solve for those
entries of the full $\boldsymbol{G}(\varepsilon)$, \textit{i.e.,}

\vspace{-0.5cm}

\begin{equation}
\overline{\boldsymbol{G}}(\varepsilon)\!=\!\left[\mathcal{I}\!-\!\overline{\boldsymbol{G}_{0}}(\varepsilon)\!\cdot\!\mathcal{V}\right]^{{\scriptscriptstyle -1}}\!\!\!\cdot\overline{\boldsymbol{G}_{0}}(\varepsilon),
\end{equation}
where $\overline{\boldsymbol{G}_{0}}(\varepsilon)$ is the clean lattice
Green's function restricted to $\Omega$, with $\mathcal{I}\!-\!\overline{\boldsymbol{G}_{0}}(\varepsilon)\!\cdot\!\mathcal{V}$
defined within the (finite-dimensional) Hilbert subspace of $\Omega$,
and\textit{ ii) }The continuation of $\boldsymbol{G}(\varepsilon)$
to the exterior of $\Omega$ is obtained via

\vspace{-0.4cm}

\begin{equation}
\boldsymbol{G}(\varepsilon)\!=\!\boldsymbol{G}_{0}(\varepsilon)\!+\!\boldsymbol{G}_{0}(\varepsilon)\!\cdot\!\overline{\mathcal{T}_{\varepsilon}}\!\cdot\!\boldsymbol{G}_{0}(\varepsilon),
\end{equation}
where

\vspace{-0.75cm}
\begin{align}
\overline{\mathcal{T}_{\varepsilon}}\! & =\!\mathcal{V}\!+\!\mathcal{V}\!\cdot\!\overline{\boldsymbol{G}}(\varepsilon)\!\cdot\!\mathcal{V}\label{eq:T_mat}\\
 & =\mathcal{V}\!\cdot\!\left[\mathcal{I}\!-\!\overline{\boldsymbol{G}}_{0}(\varepsilon)\!\cdot\!\mathcal{V}\right]^{{\scriptscriptstyle -1}}\nonumber 
\end{align}
is the projected $T$-matrix of the vacancy.

The pGF method provides access to the electronic structure of an isolated
impurity or defect embedded in an otherwise perfect infinite crystal\,\citep{Mahan1995,Buchhold2018_2}.
In the present section\textcolor{black}{, we are interested in }\textit{\textcolor{black}{i)}}\textcolor{black}{{}
the DoS change induced by a vacancy, and }\textit{\textcolor{black}{ii)}}\textcolor{black}{{}
the possibility that a WSM can host nodal bound states around the
vacancy. The emergence of zero-energy modes due to disorder is not
obvious given the absence of non-spatial symmetries in our model;
note that $\mathcal{H}_{0}$ belongs to the orthogonal Wigner-Dyson
class (class AI in the Altland-Zirnbauer ten-fold classification\,\citep{Pixley2021}).
This is to be contrasted to the well-studied case of graphene (chiral
orthogonal BDI class\,\citep{Pereira06,Ostrovski2014}), which supports
zero-energy states localized around point defects whose peculiar spectral
and transport properties have been linked to the underlying chiral
symmetry of that model\,\citep{Ferreira2015,Chiu2016}.}

One of the most readily available observable from the pGF formalism
is the change in the (extensive) DoS due to the vacancy, $\delta\nu(\varepsilon)$.
This can be evaluated by means of equation,

\vspace{-0.5cm}\textcolor{black}{
\begin{align}
\delta\nu(\varepsilon)\! & =\!\frac{1}{\pi n_{b}}\text{Im}\!\left(\text{Tr}\left[\boldsymbol{G}_{0}(\varepsilon)\!-\!\boldsymbol{G}(\varepsilon)\right]\right)\nonumber \\
 & =\!-\frac{1}{\pi n_{b}}\text{Im}\!\left(\text{Tr}\left[\mathcal{V}\!\cdot\!\left[\mathcal{I}\!-\!\boldsymbol{G}_{0}(\varepsilon)\!\cdot\!\mathcal{V}\right]^{{\scriptscriptstyle -1}}\!\!\!\cdot\left(\boldsymbol{G}_{0}(\varepsilon)\right)^{2}\right]\right)\nonumber \\
 & =\!\frac{1}{\pi n_{b}}\text{Im}\!\left(\text{tr}\left[\overline{\mathcal{T}_{\varepsilon}}\cdot\frac{d}{d\varepsilon}\boldsymbol{G}_{0}(\varepsilon)\right]\right),\label{eq:Density of States}
\end{align}
}where $\overline{\mathcal{T}_{\varepsilon}}$ is the projected $T$-matrix,
$\text{Tr}\left[\cdots\right]$ is the trace operation over all degrees
of freedom, $\text{tr}\left[\cdots\right]$ is a trace over the support
of $\mathcal{V}$, $\Omega$, and $n_{b}$ is the number of orbitals
per unit cell ($n_{b}\!\!=\!\!2$ in the WSM \textcolor{black}{model).
In the last step to obtain Eq.\,(\ref{eq:Density of States}), we
have used Eq.\,(\ref{eq:T_mat}) as well as the identity $d\boldsymbol{G}_{0}(\varepsilon)/d\varepsilon\!=\!-\left(\boldsymbol{G}_{0}(\varepsilon)\right)^{2}$.}

Next, we discuss briefly how to extract information on bound states
within the pGF framework. We start by writing the Lippmann-Schwinger
equation for a scattering state $\ket{\Psi_{\varepsilon}}$,

\vspace{-0.5cm}
\begin{equation}
\ket{\Psi_{\varepsilon}}\!=\!\ket{\Psi_{\varepsilon}^{0}}\!+\!\boldsymbol{G}_{0}(\varepsilon)\!\cdot\!\mathcal{V}\ket{\Psi_{\varepsilon}},\label{eq:LippamannSchwinger}
\end{equation}
where $\ket{\Psi_{\varepsilon}^{0}}$ is an eigenstate of the unperturbed
system, which is the parent extended state of $\ket{\Psi_{\varepsilon}}$.
In contrast, a bound state can exist without any parent eigenstate
of the clean Hamiltonian. Thus, an eigenstate bound by $\mathcal{V}$
at an energy $\varepsilon_{b}$ must be a solution of

\vspace{-0.5cm}

\begin{equation}
\ket{\Psi_{\varepsilon_{b}}}\!=\!\boldsymbol{G}_{0}(\varepsilon_{b})\!\cdot\!\mathcal{V}\ket{\Psi_{\varepsilon_{b}}}.\label{eq:LippamannSchwinger-2}
\end{equation}
Since the perturbation $\mathcal{V}$ has a finite support, one can
once again consider the projected version of Eq.\,(\ref{eq:LippamannSchwinger}),

\vspace{-0.5cm}

\begin{equation}
\ket{\xi_{\varepsilon}}\!=\!\overline{\boldsymbol{G}_{0}}(\varepsilon)\!\cdot\!\mathcal{V}\ket{\xi_{\varepsilon}},\label{eq:LippamannSchwinger-1}
\end{equation}
where $\ket{\xi_{\varepsilon_{b}}}$ is the restriction of $\ket{\Psi_{\varepsilon_{b}}}$
to the support $\Omega$. Thereby, any bound state must obey the condition,

\vspace{-0.5cm}
\begin{equation}
\left[\mathcal{I}\!-\!\overline{\boldsymbol{G}_{0}}(\varepsilon_{b})\!\cdot\!\mathcal{V}\right]\ket{\xi_{\varepsilon_{b}}^{b}}\!=\!0,\label{eq:LippamannSchwinger-1-1}
\end{equation}
which means that its projected wavefunction must belong to the kernel
of the operator, $\mathcal{I}\!-\!\overline{\boldsymbol{G}_{0}}(\varepsilon_{b})\!\cdot\!\mathcal{V}$.
Outside the support of $\mathcal{V}$ the wavefunction may be reconstructed
using

\vspace{-0.5cm}\textcolor{black}{
\begin{equation}
\Psi_{\alpha}^{b}\!\left(\mathbf{R}\right)\!=\!\braket{\mathbf{R},\alpha}{\Psi_{\varepsilon_{b}}^{b}}\!=\!\bra{\mathbf{R},\alpha}\!\boldsymbol{G}_{0}(\varepsilon_{b})\!\cdot\!\mathcal{V}\ket{\xi_{\varepsilon_{b}}^{b}},\label{eq:label}
\end{equation}
which may or may not amount to a normalizable state, depending on
the asymptotic behavior of the clean LGF. In this context, since we
are looking for zero-energy modes ($\varepsilon_{b}\!=\!0$), any
state obeying Eq.\,(\ref{eq:LippamannSchwinger-1-1}) is guaranteed
to be square-normalizable in 3D space with an algebraic tail $\propto r^{{\scriptscriptstyle -2}}$.
The latter is the long-distance behavior of $\boldsymbol{G}_{0}(0,\boldsymbol{\Delta R})$,
as obtained in the continuum limit.}

\vspace{-0.45cm}

\subsubsection{\label{subsec:Lattice-Vacancies-in}Full-Vacancies in a Weyl Semimetal}

\vspace{-0.2cm}

We now apply the general formalism described above to the case of
a \textit{full-vacancy} where both orbitals are removed from a particular
lattice cell. To perform the calculation, it comes in handy to order
the sites of $\Omega$ as $\left\{ \boldsymbol{0},a\hat{x},a\hat{y},a\hat{z},-a\hat{x},-a\hat{y},-a\hat{z}\right\} $.
With this ordering, one obtains $\mathcal{V}$ as the matrix

\vspace{-0.5cm}
\begin{equation}
\mathcal{V}\!=\!\frac{\hbar v}{2a}\left[\begin{array}{ccc}
\mathbb{O}_{{\scriptscriptstyle 2\!\times\!2}} & \!\!\!-\boldsymbol{\sigma} & \boldsymbol{\sigma}\\
\boldsymbol{\sigma}^{T} & \mathbb{O}_{{\scriptscriptstyle 6\!\times\!6}} & \mathbb{O}_{{\scriptscriptstyle 6\!\times\!6}}\\
\!\!\!-\boldsymbol{\sigma}^{T} & \mathbb{O}_{{\scriptscriptstyle 6\!\times\!6}} & \mathbb{O}_{{\scriptscriptstyle 6\!\times\!6}}
\end{array}\right],
\end{equation}
within the projected subspace, while, by exploiting the symmetries
of $\mathbf{G}_{0}$ {[}Eqs.\,\ref{eq: symmetry 1} and \ref{eq: symmetry 2}{]},
we are able to further express the pGF as

\vspace{-0.5cm}

\begin{equation}
\overline{\boldsymbol{G}_{0}}(0)\!=\!\frac{a}{i\hbar v}\!\!\left[\,\,\begin{array}{ccc}
\mathbb{O}_{{\scriptscriptstyle 2\!\times\!2}} & g_{0}\boldsymbol{\sigma} & \!\!\!\!-g_{0}\boldsymbol{\sigma}\\
\!\!\!\!-g_{0}\boldsymbol{\sigma}^{T} & \mathbb{O}_{{\scriptscriptstyle 6\!\times\!6}} & \mathbb{O}_{{\scriptscriptstyle 6\!\times\!6}}\\
g_{0}\boldsymbol{\sigma}^{T} & \mathbb{O}_{{\scriptscriptstyle 6\!\times\!6}} & \mathbb{O}_{{\scriptscriptstyle 6\!\times\!6}}
\end{array}\,\right],
\end{equation}
wher\textcolor{black}{e $g_{0}\!=\!\mathcal{I}_{i0^{+}}^{x}\!\left(1,0,0\right)$.
}These simple matrices can then be used to build the operator $\mathcal{I}\!-\!\overline{\boldsymbol{G}_{0}}(0)\!\cdot\!\mathcal{V}$,
whose determinant takes the remarkably simple form,

\vspace{-0.5cm}
\begin{equation}
\det\left(\mathcal{I}\!-\!\overline{\boldsymbol{G}_{0}}(0)\!\cdot\!\mathcal{V}\right)\!=\!\left(i\!-\!3g_{0}\right)^{4}.\label{eq:DeterminantWeyl}
\end{equation}
Equation\,(\ref{eq:DeterminantWeyl}) has a clear physical interpretation:
a \textit{four-fold} \textit{degenerate} root appears for $g_{0}\!=\!\nicefrac{i}{3}$,
corresponding to an extra pair of non-trivial bound states extending
into the lattice\textcolor{black}{. For consistency, $\det\left(\mathcal{I}\!-\!\overline{\boldsymbol{G}_{0}}(0)\!\cdot\!\mathcal{V}\right)$
would have to display a }\textit{\textcolor{black}{two-fold degenerate}}\textcolor{black}{{}
root, corresponding to the subspace of both orbitals to be removed
from the lattice. Surprisingly, the degeneracy appears doubled here,
which indicates that two nodal bound states must exist as a mathematical
property of a full-vacancy in our lattice model.} In addition, a full
diagonalization of $\mathcal{I}\!-\!\overline{\boldsymbol{G}_{0}}(0)\!\cdot\!\mathcal{V}$
yields the following projected wavefunctions for these states:

\vspace{-0.5cm}

\begin{subequations}
\begin{align}
\ket{\xi_{1}^{b}} & =\frac{1}{\sqrt{6}}\left[\ket{\hat{\mathbf{x}},1}-i\ket{\hat{\mathbf{y}},1}-\ket{\hat{\mathbf{z}},2}\right.\\
 & \qquad-\left.\ket{-\hat{\mathbf{x}},1}+i\ket{-\hat{\mathbf{y}},1}+\ket{-\hat{\mathbf{z}},2}\right],\nonumber 
\end{align}

\vspace{-0.8cm}

\begin{align}
\ket{\xi_{2}^{b}} & =\frac{1}{\sqrt{6}}\left[\ket{\hat{\mathbf{x}},2}\right.+i\ket{\hat{\mathbf{y}},2}+\ket{\hat{\mathbf{z}},1}\\
 & \qquad-\left.\ket{-\hat{\mathbf{x}},2}-i\ket{-\hat{\mathbf{y}},2}-\ket{-\hat{\mathbf{z}},1}\right],\nonumber 
\end{align}
\end{subequations}
where $\ket{\mathbf{R},\alpha}$ are the local Wannier states (here,
$\alpha$ indexes the orbital). Upon a reconstruction, these states
have the following real-space wavefunction outside $\Omega$:

\vspace{-0.4cm}

\begin{subequations}
\begin{align}
\Psi_{1}^{b}\!\left(\mathbf{R}\right)\! & =\!\frac{it\sqrt{6}}{2}\left[\begin{array}{c}
G_{0}^{12}\!\left(0,\mathbf{R}\right)\\
G_{0}^{22}\!\left(0,\mathbf{R}\right)
\end{array}\right]\\
\Psi_{2}^{b}\!\left(\mathbf{R}\right)\! & =\!\frac{it\sqrt{6}}{2}\left[\begin{array}{c}
G_{0}^{11}\!\left(0,\mathbf{R}\right)\\
G_{0}^{21}\!\left(0,\mathbf{R}\right)
\end{array}\right]\!,
\end{align}
\end{subequations}

\noindent where $G_{0}^{\alpha\beta}$\textcolor{black}{{} are spinor
components of $\mathbf{G}_{0}$.}

\textcolor{black}{In a similar manner, we can determine the DoS change
caused by a vacancy defect. To do this, we require the clean pGF at
all energies, which has the following matrix structure,}

\textcolor{black}{\vspace{-0.3cm}}

\textcolor{black}{
\begin{equation}
\overline{\boldsymbol{G}_{0}}(\varepsilon)\!=\!\frac{a}{i\hbar v}\!\!\left[\,\,\begin{array}{ccc}
f_{\varepsilon}\mathbb{I}_{{\scriptscriptstyle 2\!\times\!2}} & g_{\varepsilon}\boldsymbol{\sigma} & \!\!\!\!-g_{\varepsilon}\boldsymbol{\sigma}\\
\!\!\!\!-g_{\varepsilon}\boldsymbol{\sigma}^{T} & \,\,f_{\varepsilon}\mathbb{I}_{{\scriptscriptstyle 6\!\times\!6}} & h_{\varepsilon}\mathbb{I}_{{\scriptscriptstyle 6\!\times\!6}}\\
g_{\varepsilon}\boldsymbol{\sigma}^{T} & \,\,h_{\varepsilon}\mathbb{I}_{{\scriptscriptstyle 6\!\times\!6}} & f_{\varepsilon}\mathbb{I}_{{\scriptscriptstyle 6\!\times\!6}}
\end{array}\,\right]\!\!,\label{eq:pGF_Clean}
\end{equation}
as imposed by the aforementioned symmetries of our WSM model. In Eq.\,(\ref{eq:pGF_Clean}),
$f_{\varepsilon}\!=\!\mathcal{I}_{\varepsilon}^{0}\left(0,0,0\right)$,
$g_{\varepsilon}\!=\!\mathcal{I}_{\varepsilon}^{x}\left(1,0,0\right)$
and $h(\varepsilon)\!=\!\mathcal{I}_{\varepsilon}^{0}\left(2,0,0\right)$
are dimensionless functions of the energy variable. Having the pGF,
we can now build the $T$-matrix of an isolated full-vacancy and employ
Eq.\,(\ref{eq:Density of States}) to obtain,}

\begin{figure}[t]
\begin{centering}
\hspace{-0.4cm}\includegraphics[scale=0.65]{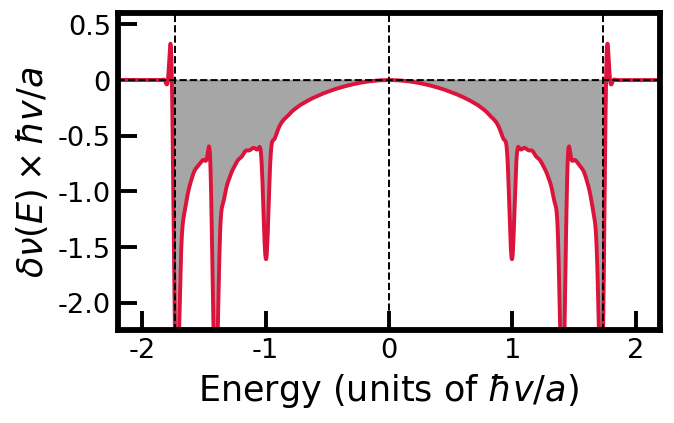}
\par\end{centering}
\begin{centering}
\vspace{-0.5cm}
\par\end{centering}
\caption{\label{fig:DoSCorrection}DoS correction due \textcolor{black}{to
a single (full) vacancy in an infinite lattice (calculated with numerical
differentiation of the lattice GF). The unperturbed bandwidth of the
model is marked by the outermost vertical dashed lines and, unlike
the conventional case, integrating $\delta\nu(E)$ over the entire
band yields $-2$ instead of zero}.}

\vspace{-0.5cm}
\end{figure}

\vspace{-0.4cm}
\begin{equation}
\!\!\!\delta\nu(\varepsilon)\!=\!\frac{3a}{\pi\hbar v}\text{Im}\!\!\left[\!\frac{f_{\varepsilon}\left(h_{\varepsilon}^{\prime}\!-\!2f_{\varepsilon}^{\prime}\right)\!+\!f_{\varepsilon}^{\prime}h_{\varepsilon}\!-\!4g_{\varepsilon}^{\prime}\left(i\!+\!3g_{\varepsilon}\right)}{3f_{\varepsilon}^{2}\!-\!3f_{\varepsilon}h_{\varepsilon}\!+\!2\left(i\!+\!3g_{\varepsilon}\right)^{2}}\!\right]\!\!.\label{eq:CorrectionDoS_Weyl}
\end{equation}
The functions $f_{\varepsilon}$, $g_{\varepsilon}$ and $h_{\varepsilon}$
and their derivatives were calculated nu\textcolor{black}{merically
(see Appendix\,\ref{sec:The-Clean-Lattice}) and the resulting DoS
is shown in Fig.\,\ref{fig:DoSCorrection}. As expected, a single
(full) vacancy causes a negative correction to the DoS across the
entire band, which is consistent with an overall transfer of spectral
weight to the} emergent bound states around the vacant site. The integral
of this curve is exactly $-2$, as the number of continuum states
is not conserved, \textit{i.e.}, two states (per orbital) appear as
a vacancy bound-states and other two are removed from the Hilbert
space.

\vspace{-0.4cm}

\subsubsection{Half-Vacancies in Weyl Semimetals}

\vspace{-0.2cm}

We now consider the effects of an isolated \textit{half-vacancy} in
which only one orbital is removed from each cell. If the orbital $1$
is vacant, the perturbation reads

\vspace{-0.5cm}
\begin{align}
\mathcal{V}_{\!1}\! & =\!-\frac{\hbar v}{2a}\left[\left(\psi_{\boldsymbol{0}}^{1}\right)^{\dagger}\!\!\psi_{a\hat{x}}^{2}\!-\!i\left(\psi_{\boldsymbol{0}}^{1}\right)^{\dagger}\!\!\psi_{a\hat{y}}^{2}+\!\left(\psi_{\boldsymbol{0}}^{1}\right)^{\dagger}\!\!\psi_{a\hat{z}}^{1}\!\right.\\
 & \left.-\left(\psi_{\boldsymbol{0}}^{1}\right)^{\dagger}\!\!\psi_{-a\hat{x}}^{2}\!\!+\!i\left(\psi_{\boldsymbol{0}}^{1}\right)^{\dagger}\!\!\psi_{-a\hat{y}}^{2}\!\!-\!\left(\psi_{\boldsymbol{0}}^{1}\right)^{\dagger}\!\!\psi_{-a\hat{z}}^{1}\!-\!\text{h.c.}\right],\nonumber 
\end{align}
while in the opposite case, it reads

\vspace{-0.5cm}

\begin{align}
\mathcal{V}_{\!2}\! & =\!-\frac{\hbar v}{2a}\left[\left(\psi_{\boldsymbol{0}}^{2}\right)^{\dagger}\!\!\psi_{a\hat{x}}^{1}\!+\!i\left(\psi_{\boldsymbol{0}}^{2}\right)^{\dagger}\!\!\psi_{a\hat{y}}^{1}-\!\left(\psi_{\boldsymbol{0}}^{2}\right)^{\dagger}\!\!\psi_{a\hat{z}}^{2}\!\right.\\
 & \left.-\left(\psi_{\boldsymbol{0}}^{2}\right)^{\dagger}\!\!\psi_{-a\hat{x}}^{1}\!\!-\!i\left(\psi_{\boldsymbol{0}}^{2}\right)^{\dagger}\!\!\psi_{-a\hat{y}}^{1}\!+\!\left(\psi_{\boldsymbol{0}}^{2}\right)^{\dagger}\!\!\psi_{-a\hat{z}}^{2}\!-\!\text{h.c.}\right].\nonumber 
\end{align}
In either situation, the presence of bound-states and DoS deformations
can be investigated along the same lines of the \textit{full-vacancy},
the only difference being the projected perturbation matrix. More
specifically, we write

\vspace{-0.5cm}

\begin{equation}
\mathcal{V}_{1/2}\!=\!\frac{\hbar v}{2a}\left[\begin{array}{ccc}
\mathbb{O}_{{\scriptscriptstyle 2\!\times\!2}} & \!\!\!-\!\boldsymbol{\Sigma}_{1/2} & \boldsymbol{\Sigma}_{1/2}\\
\boldsymbol{\Sigma}_{1/2}^{T} & \mathbb{O}_{{\scriptscriptstyle 6\!\times\!6}} & \mathbb{O}_{{\scriptscriptstyle 6\!\times\!6}}\\
\!\!\!-\!\boldsymbol{\Sigma}_{1/2}^{T} & \mathbb{O}_{{\scriptscriptstyle 6\!\times\!6}} & \mathbb{O}_{{\scriptscriptstyle 6\!\times\!6}}
\end{array}\right],
\end{equation}
where the $(2\!\times\!6)$-dimensional $\boldsymbol{\Sigma}$--matrices
are

\vspace{-0.5cm}

\begin{subequations}
\begin{align}
\boldsymbol{\Sigma}_{\text{u}}\! & =\!\left[\begin{array}{cccccc}
0 & 1 & 0 & -i & 1 & 0\\
0 & 0 & 0 & 0 & 0 & 0
\end{array}\right]\,,\\
\boldsymbol{\Sigma}_{\text{l}}\! & =\!\left[\begin{array}{cccccc}
0 & 0 & 0 & 0 & 0 & 0\\
1 & 0 & i & 0 & 0 & -1
\end{array}\!\right].
\end{align}
\end{subequations}

\noindent The clean pGF is exactly the same as in Eq.\,(\ref{eq:pGF_Clean}),
and therefore

\vspace{-0.5cm}

\begin{equation}
\det\left(\mathcal{I}\!-\!\overline{\boldsymbol{G}_{0}}(0)\!\cdot\!\mathcal{V}_{1/2}\right)\!=\!-\left(i\!-\!3g_{0}\right)^{2}
\end{equation}
which has a double root for $g_{0}\!=\!\nicefrac{i}{3}$. The diagonalization
of this operator \textcolor{black}{confirms that its null-space is}
a two-dimensional subspace generated by the removed orbital, plus
a non-trivial bound state that surrounds the vacant site. Similarly,
the correction to the DoS is exactly the same as Eq.\,(\ref{eq:CorrectionDoS_Weyl})
but with an added factor of $\nicefrac{1}{2}$.\textcolor{black}{{}
The similar behavior between }\textit{\textcolor{black}{half-}}\textcolor{black}{{}
and }\textit{\textcolor{black}{full-vacancies}}\textcolor{black}{{}
could have been anticipated by looking at a }\textit{\textcolor{black}{full-vacanc}}\textcolor{black}{y
as a pair of }\textit{\textcolor{black}{half-vacancies}}\textcolor{black}{{}
placed within the same unit cell. In the pGF formalism, these correspond
to local perturbations ($\mathcal{V}_{1}$ and $\mathcal{V}_{2}$)
that act in disjoint Hilbert subspaces not connected by the clean
lattice propagator. This prevents the two }\textit{\textcolor{black}{half-vacancies}}\textcolor{black}{{}
from hybridizing and their resulting effects in the spectrum will
be simply cumulative. For this reason, we focus exclusively on }\textit{\textcolor{black}{full-vacancies}}\textcolor{black}{{}
in the remainder of this paper. }

\vspace{-0.3cm}

\section{\label{sec:Robustness to Disorder}Microscopic Robustness of the
Vacancy Bound States}

After establishing the existence of vacancy-induced nodal bound states,
we move on to assess their robustness against additional disorder
sources. For that, we model the additional disordered landscape as
an uncorrelated scalar potential, $V_{d}(\boldsymbol{R})$. The Hamiltonian
now reads 
\begin{figure}[t]
\begin{centering}
\hspace{-0.1cm}\includegraphics[scale=0.4]{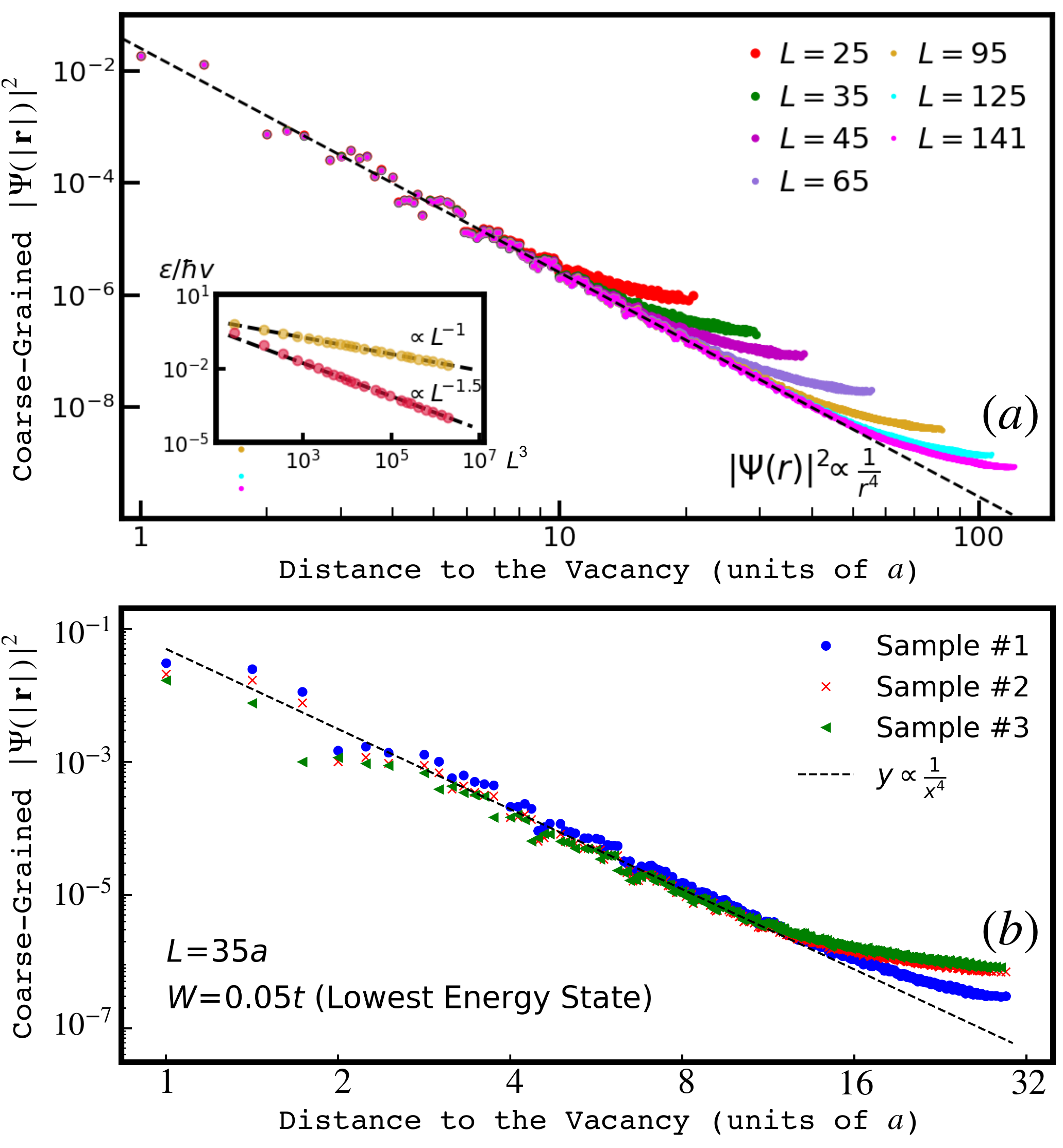}
\par\end{centering}
\vspace{-0.3cm}

\caption{\label{fig:Wavefunction}Probability density, $\protect\abs{\Psi\!\left(\mathbf{R}\right)}^{2}\!\!=\!\protect\abs{\Psi_{1}\!\left(\mathbf{R}\right)}^{2}\!+\!\protect\abs{\Psi_{2}\!\left(\mathbf{R}\right)}^{2}$,
of the eigenstate closest to zero averaged over spherical shells of
width $a$ cen\textcolor{black}{tered on the vacancy. (a) Single central
vacancy with no additional disorder. Inset: The eigenvalues obtained
near $E\!=\!0$ for finite samples come in two flavors: those which
scale to zero energy as $L^{{\scriptscriptstyle -1}}$ (orange), akin
to the predicted mean-level spacing scaling around a clean Weyl node,
and eigenvalues which scale faster {[}as $\propto L^{{\scriptscriptstyle -3/2}}${]}
due to the more localized character of the corresponding eigenstate
(magenta). (b) Vacancy within three random disordered environments.}}

\vspace{-0.5cm}
\end{figure}

\vspace{-0.3cm}
\begin{equation}
\mathcal{H}_{d}\!=\!\mathcal{H}_{0}\!+\!\mathcal{V}\!+\!\sum_{\mathbf{R}\in\mathcal{L}}V_{d}(\boldsymbol{R})\Psi_{\boldsymbol{R}}^{\dagger}\!\cdot\!\Psi_{\boldsymbol{R}}.\label{eq:Ham_Dis}
\end{equation}

To tackle this problem, we numerically diagonalize $\mathcal{H}_{d}$
around $E\!=\!0$ using the $\texttt{scipy}$ implementation of the
implicitly restarted Lanczos method\,\citep{Lanczos50,paige1980,Lehoucq97}.
Since the method converges better to non-clustered eigenpairs in the
borders of the spectrum, we apply it to $\mathcal{H}_{d}^{2}$ instead
and restrict the analysis to low-lying eigenstates. Additionally,
we consider cubic samples of side $L$, with a sing\textcolor{black}{le
(full) vacancy at the center of each sample, supplemented by fixed
phase-twisted boundaries that open a finite-size gap ($\Delta_{f}\!\propto\!L^{{\scriptscriptstyle -1}}$)
in the spectrum of extended states. The nodal bound states will lie
inside this finite-size gap, as they are weakly affected by the boundary
conditions.}

\textcolor{black}{In Fig.\,\ref{fig:Wavefunction}\,a, we represent
the radial distribution of the vacancy bound states for different
simulation sizes, in the absence of any additional disorder. The results
confirm the predicted zero-energy states with tails decaying as $r^{{\scriptscriptstyle -2}}$.
We note that the degeneracy of these states gets slightly lifted by
the boundary conditions but the corresponding eigenvalues still tends
to zero as $L^{{\scriptscriptstyle -\frac{3}{2}}}$, }\textit{\textcolor{black}{i.e.,}}\textcolor{black}{{}
faster than $\Delta_{f}$. Furthermore, the vacancy defect perturbs
slightly the extended states (now scattering states), which further
expands the finite-size gap\,}\footnote{\textcolor{black}{This difference is too small to be seen in the inset
of Fig.\,\ref{fig:Wavefunction} and does not affect the asymptotic
scaling of $\Delta_{f}$ with the system size.}}\textcolor{black}{.}

\textcolor{black}{Next, we present an identical analysis with the
vacancy's surroundings endowed with an uncorrelated random scalar
potential uniformly drawn from $[-\nicefrac{W}{2},\nicefrac{W}{2}]$.
In principle, this alteration dresses the LGF of the clean model,
thus }\textit{\textcolor{black}{destroying most model-specific symmetries.}}\textcolor{black}{{}
In }Fig.\,\ref{fig:Wavefunction}\,b, we present the radial wavefunctions
of the two eigenstates closest to zero energy obtained from the diagonalization
of three randomly generated disorder configurations. In all three
cases, the states feature the same normalizable power-law tail found
in the clean case, indicating that the vacancy bound states are indeed
robust.

To further understand the effects of an uncorrelated disorder landscape,
we diagonalized $10^{{\scriptscriptstyle 4}}$ systems with randomly
generated disorder samples around a single central vacancy, focusing
on determining the four eigenpairs whose energies are the closest
to the node. In addition to the eigenenergies, we used the eigenfunctions
to determine the inverse participation ratio (IPR),

\vspace{-0.6cm}

\begin{equation}
\text{IPR}_{\Psi}\!=\frac{\sum_{\mathbf{R}}\left(\abs{\Psi_{\mathbf{R}}^{1}}^{2}\!\!+\!\abs{\Psi_{\mathbf{R}}^{2}}^{2}\right)^{2}}{\sum_{\mathbf{R}}\abs{\Psi_{\mathbf{R}}^{1}}^{2}\!\!+\!\abs{\Psi_{\mathbf{R}}^{2}}^{2}},\label{eq:IPR}
\end{equation}
a simple quantity that allows one to distinguish well localized, $\text{IPR}\!\sim\!\mathcal{O}(1)$,
from delocalized states, for which $\text{IPR}\!\sim\!\mathcal{O}(L^{{\scriptscriptstyle -3}})$.
In Fig.\,\ref{fig:DisorderHistograms}\,a and \ref{fig:DisorderHistograms}\,b,
we show histograms of the eigenenergies for three system sizes and
two disorder strengths, using twisted boundary conditions with a fixed
twist angle of $\nicefrac{\pi}{3}$ in all directions. These histograms
borne out two well-separated clusters formed by: \textit{(a)} the
two eigenstates closest to zero energy which are broadened by disorder
around $E\!=\!0$, but remain firmly inside the finite-size gap, and
\textit{(b)} the ones corresponding to the largest eigenvalues, being\textcolor{black}{{}
Bloch states that get} perturbatively shifted towards the node and
broadened by disorder\,\citep{Pixley2021}. As confirmed by the corresponding
IPRs, the \textit{(a)}-class states are strongly localized states
which are still bound to the central vacancy, while the \textit{(b)}-class
are disorder-dressed extended Weyl states.

\vspace{-0.4cm}

\section{\label{sec:Finite-Vacancy-Concentrations}Quantum-Interference and
Finite Concentration Effects}

\noindent The previous results established that a singl\textcolor{black}{e
(full) vacancy defect gives rise to a pair of zero-energy bound states
with power-law-localized wavefunctions. We now discuss the effect
of coherent multiple scattering events in realistic systems\,\citep{Besara2016}
which have a finite (nonzero) concentration of point defec}ts. The
main question we ask here is whether the essential IPR features of
zero-energy states survive the unavoidable inter-vacancy hybridization
effects. 

\begin{figure}[H]
\begin{centering}
\hspace{-0.2cm}\includegraphics[scale=0.26]{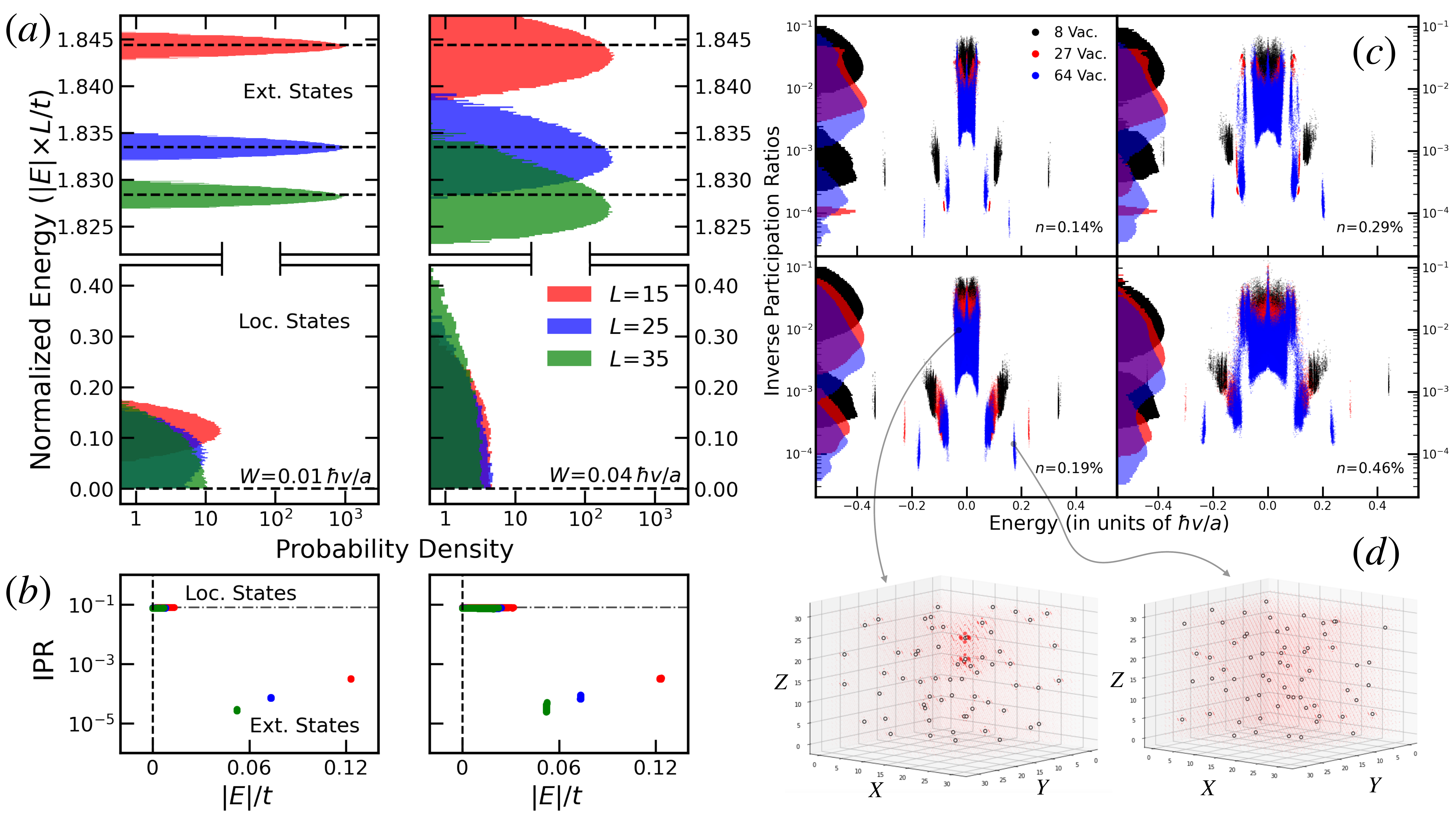}
\par\end{centering}
\vspace{-0.2cm}

\caption{\label{fig:DisorderHistograms}\textbf{(a)} Histograms of the four
energies closest to the Weyl node for a vacancy surrounded by a random
Anderson potential of strength $W\!=\!0.01\hbar v/a$ (right) and
$W\!=\!0.04\hbar v/a$ (left), represented for systems with different
linear sizes {[}$L=15$ (red), $L=25$ (blue) and $L=35$ (red){]}.
Black dashed lines indicate the clean energy levels and, thus, the
corresponding finite size gaps. \textbf{(b)} Scatter plot of the energies
and IPRs of the calculated eigenpairs. The two eigenstates closest
to zero energy are very localized in real-space, while the remaining
pair is clearly delocalized. \textbf{(c) }Energy-IPR scatter plot
for the eigenstates closest to the node obtained for $25000$ samples
of a WSM with four concentrations of randomly placed full-vacancies.
\textcolor{black}{Histograms of the IPRs are presented for each total
number of vacancies along the vertical axis. Different colors label
different vacancy numbers (here, the simulated cell size adjusted
to guarantee that $n$ is constant in all 4 cases presented). }\textbf{(d)}
Bubble chart representation of the (squared) wavefunctions for two
eigenstates picked at random from the indicated regions: a heavily
localized state around a few vacancies, (left) and an extended state
(right).}
\end{figure}

Our starting approach to this problem is based upon the exact diagonalization
of small systems. We consider WSM lattices with linear sizes up to
$L=35$ and a concentration ($n$) of randomly placed \textit{full-vacancies}.
By means of the twisted boundary conditions, we open a finite-size
gap that separates nodal bound states from extended ones. The $2nL^{3}\!+\!4$
eigenpairs\,\footnote{We expect each vacancy to generate a pair of bound states, making
this number sufficiently large to capture most states within the finite-size
gap.} closest to $E\!=\!0$ are then extracted using LD. In Fig.\,\ref{fig:DisorderHistograms}\,c
we represent a scatter plot of the energies and corresponding IPRs
of every eigenpair determined for $2500$ random arrangements of vacancies
with concentrations ranging from $0.1\%$ to $1\%$ (per unit cell).
The results clearly demonstrate that, in spite of the proximity between
vacancies, the system still features a large number of high-IPR eigenstates
which are flanked by a region of extended states. Such a physical
interpretation is clear from Fig.\,\,\ref{fig:DisorderHistograms}\,d,
where a 3D bubble chart of $\abs{\Psi(\mathbf{R})}^{2}$ is depicted
for two eigenstates randomly chosen from each of the regions.

\onecolumngrid

\twocolumngrid

\subsection{\label{subsec:Average-Density-of}Average Density of States}

\vspace{-0.1cm}

The LD study gives a qualitative picture of the structure of the eigenstates
surrounding a Weyl node, but its utility can be severely limited by
loss of spectral resolution, the finite number of eigenstates that
are accessible and, lastly, the attainable system sizes. Therefore,
we now complement the LD analysis with full-spectral simulations of
the DoS of large systems by means of the \textit{kernel polynomial
method} (KPM)\,\citep{Weise2006}. As a first step, we present results
on the ensemble-averaged DoS for a large system with a linear size
$L\!=\!512$. This observable gives us information on how the spectrum
is modified by inter-vacancy hybridization effects, yielding a numerically
exact picture of the vacancy-induced resonances\,\citep{Pires2021,Buchhold2018_2}
around the node. \textcolor{black}{The KPM calculations are carried
out with domain decomposition and a stochastic trace evaluation techniques
as implemented in KITE\,\citep{Joao2020}. The calculation employs
$M\!=\!2^{16}$ Chebyshev moments (corresponding to a spectral resolution
$\eta\!=\!10^{-4}\hbar v/a$), a Jackson kernel, and a sufficiently
large number of random vectors to yield highly converged results.
Finally, the results are averaged over random twisted boundary conditions
which eliminates the finite size mean-level spacing.}

\begin{figure}[t]
\vspace{-0.2cm}
\begin{centering}
\hspace{-0.2cm}\includegraphics[scale=0.23]{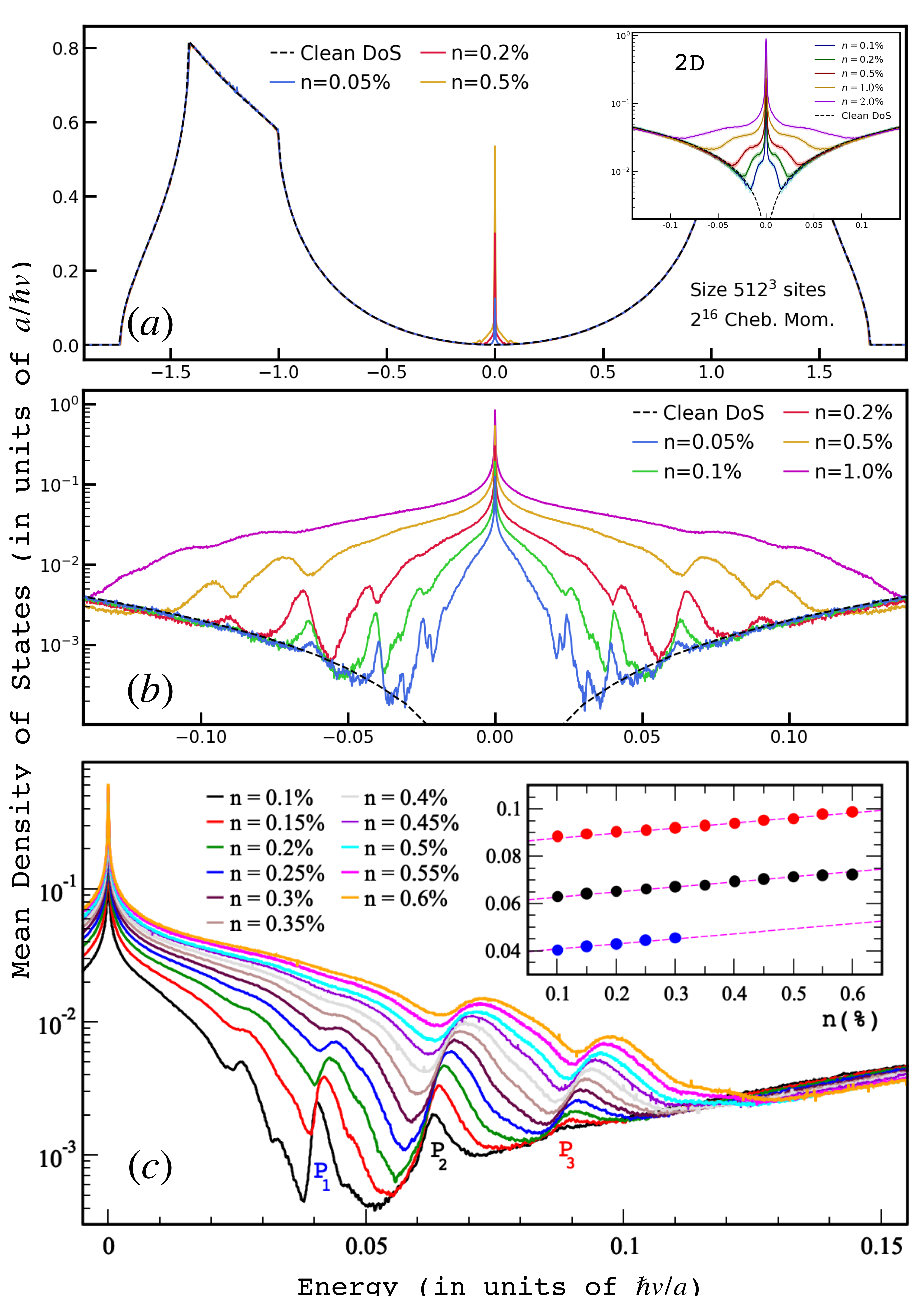}
\par\end{centering}
\vspace{-0.35cm}

\caption{\label{fig:KPM_withConcentration}DoS of a WSM lattice of linear size
$L\!=\!512$ for selected defect concentrations. \textbf{(a)} Bird's-eye
view of the DoS. The inset shows the two-dimensional case for comparison.\textbf{
(b)} Close-up of average DoS around $\varepsilon\!=\!0$. \textbf{(c)}
Subsidiary peaks in the DoS at low vacancy concentration. Inset: Evolution
of the the peak height ($P_{1},P_{2}\text{ and }P_{3}$) with $n$.}

\vspace{-0.5cm}
\end{figure}

The average DoS obtained through the KPM is shown in Fig.\,\ref{fig:KPM_withConcentration}.
These high-resolution results disclose a prominent spectral enhancement
in and around the node which indicates that, unlike what happens for
ordinary on-site disorder\,\citep{Nandkishore2014,Pixley2015,Bera2016,Pixley2016,Ziegler2018,Buchhold2018,Wilson2020,Pires2021,Pixley2021,Buchhold2018_2},
the DoS at $E\!=\!0$ gets quickly lifted to a large value as $n$
increases. This pronounced effect is consistent with the presence
of robust nodal bound states, and validates the conclusions of Sec.\,\ref{sec:Finite-Vacancy-Concentrations}.
Moreover, as the central DoS peak grows in height with increasing
$n$, a much wider symmetrical profile begins to emerge at its base.
As shown in Appendix\,\ref{sec:Additional-Numerics}, the integral
of this DoS correction is proportional to $n$, which indicates that
inter-vacancy hybridization is simply turning the bound states of
isolated (full) vacancies into scattering resonances within the continuum.
In Fig.\,\ref{fig:KPM_withConcentration}\,b, a closeup of this
structure is shown, revealing a finer\textit{ comb-like} structure
of subsidiary peaks (sharp scattering resonances) around the node
for moderate defect concentrations ($n\lesssim1\%$). These peaks
are more visible in Fig.\,\ref{fig:KPM_withConcentration}\,c, where
their displacement as a function of $n$ is also shown. The modulated
structure in the DoS reported in this work is a unique feature of
3D WSMs, which is absent in the analogous two-dimensional problem\,\citep{Pereira06,Ferreira2015}
(see inset to Fig.\,\ref{fig:KPM_withConcentration}\,a). The subsidiary
peaks in the DoS are robust to an additional weakly disordered potential,
as is discussed in Appendix\,\ref{sec:Additional-Numerics}. 

\vspace{-0.2cm}

\section{\label{sec:Conclusion-and-Outlook}Conclusion and Outlook}

\vspace{-0.1cm}

A combination of exact diagonalization and large-scale spectral methods
allow us to resolve the impact of point defects on the real-space
electronic structure of 3D $\mathcal{T}$-symmetric Weyl semimetals.
Our results for a lattice WSM model show that dilute concentrations
of vacancies, a common crystal imperfection, have a strong impact
on the electronic properties in stark contrast with uncorrelated on-site
disorder models\,\citep{Nandkishore2014,Pixley2015,Syzranov2015,Syzranov2016,Bera2016,Buchhold2018,Buchhold2018_2,Syzranov2018,Ziegler2018,Wilson2020,Pires2021,Pixley2021,Pixley2016}
which have been found to produce a minute effect on the low-energy
properties of Weyl systems. In fact, random vacancies were shown to
efficiently lift the nodal DoS thereby destabilizing the semi-metallic
phase even at very low concentrations. Moreover, we have also demonstrated
that quantum-interference effects between vacancies can yield a peculiar
modulated energy dependence of electronic observables which has no
analogue in two-dimensional Dirac systems\,\citep{Fradkin1986,Ostrovski2014,Ferreira2015}.
While the average DoS displays a comb of subsidiary resonance peaks
at finite energies, \textcolor{black}{we show in a companion paper\,\citep{Pires2022}
that} the bulk dc-conductivity mirrors this behavior through a series
of sudden dips as the Fermi level is varied. Therefore, upon tuning
the carrier density in real samples\,\citep{Nishihaya2018} {[}or
even the defect concentration\textit{ }using H/He\,\citep{Zhang2019}
or light-ion irradiation\,\citep{Fu2020}{]}, we predict that bulk
transport measurements will allow the observation of interesting signatures
of native point defects.\textcolor{black}{{} These are expected to assume
chief importance in WSMs of the TaAs family, whose concentration of
point defects in high-quality crystals grown by chemical vapor transport
is experimentally known to be large\,\citep{Besara2016}.} At last,
in Ref.\,\citep{Pires2022} the authors present a thorough investigation
of physical consequences from these novel effects of vacancy disorder
in electronic structure of the Weyl nodes, with a particular focus
on experimentally accessible signatures from standard transport and
optical response measurements. \textcolor{black}{The optical signatures
have a particular practical importance, as they do not rely on an
external control over the system's Fermi energy.}

\vspace{-0.4cm}
\begin{acknowledgments}
J.P.S.P., S.M.J, B.A. and J.M.V.P.L acknowledge support from the Portuguese
Foundation for Science and Technology (FCT) within the Strategic Funding
UIDB/04650/2020, and through projects No.\,POCI-01-0145-FEDER-028887
(J.P.S.P., S.M.J and J.M.V.P.L) and No.\,CEECIND/02936/2017\,(B.A.).
J.P.S.P. and S.M.J are funded by FCT grants No.\,PD/BD/142774/2018
and PD/BD/142798/2018, respectively. A.F. acknowledges support from
the Royal Society (London) through a Royal Society University Research
Fellowship. The large-scale calculations were undertaken on the HPC\textcolor{black}{{}
Viking Cluster of the University of York. We thank J. Dieplinger,
J.M.B Lopes dos Santos and A. Altland for fruitful discussions. We
also thank the anonymous referees for providing valuable feedback.}
\end{acknowledgments}

\vspace{-0.4cm}

\appendix

\section{\label{sec:The-Clean-Lattice}Lattice Green Function}

\noindent Here, we outline the semi-analytical method employed to
obtain the LGF with arbitrary spectral resolution. As indicated in
the main text, Eqs.\,(\ref{eq:GreenFunctionClean})-(\ref{eq:G0Structure}),
the clean LGF is a $2\!\times\!2$ matrix that can be written as

\vspace{-0.5cm}

\begin{equation}
\boldsymbol{G}_{0}(\varepsilon,\!\boldsymbol{\Delta R})\!=\!\mathcal{I}_{\varepsilon}^{0}\!\!\left(\boldsymbol{\Delta R}\right)\!-\!\!\!\!\sum_{j=x,y,z}\!\!\!\!i\sigma^{j}\mathcal{I}_{\varepsilon}^{j}\!\!\left(\boldsymbol{\Delta R}\right),\label{eq:G0Structure-1}
\end{equation}
with $\mathcal{I}_{\varepsilon}^{0,x,y,z}\left(n_{x},n_{y},n_{z}\right)$
being four position- and energy-dependent integrals over the cubic
fBZ. These integrals are given by

\vspace{-0.5cm}

\begin{subequations}
\begin{align}
\mathcal{I}_{\varepsilon}^{0}\!\left(\!\boldsymbol{\Delta R}\right)\! & =\!\!\int_{\text{[-\textgreek{p},\textgreek{p}]}^{3}}\!\!\!\!\!\!\!\!\!d^{{\scriptscriptstyle \text{(\!3\!)}}}\mathbf{k}\frac{\varepsilon\,e^{ik_{x}n_{x}}e^{ik_{y}n_{y}}e^{ik_{z}n_{z}}}{8\pi^{3}\left(\varepsilon^{2}+\abs{\boldsymbol{\sin k}}^{2}\right)}\label{eq:Integrals-1}\\
\mathcal{I}_{\varepsilon}^{x}\!\left(\!\boldsymbol{\Delta R}\right)\! & =\!\!\!\int_{\text{[-\textgreek{p},\textgreek{p}]}^{3}}\!\!\!\!\!\!\!\!\!d^{{\scriptscriptstyle \text{(\!3\!)}}}\mathbf{k}\frac{\sin\!k_{x}e^{ik_{x}n_{x}}e^{ik_{y}n_{y}}e^{ik_{z}n_{z}}}{8\pi^{3}\left(\varepsilon^{2}+\abs{\boldsymbol{\sin k}}^{2}\right)},
\end{align}
\end{subequations}

\noindent For a single-vacancy calculation in the lattice WSM, the
finite support of the perturbation dictates that only three of these
integrals are required, namely:

\vspace{-0.5cm}

\noindent 
\begin{subequations}
\noindent 
\begin{align}
\mathcal{I}_{\varepsilon}^{0}\!\left(0,\!0,\!0\right)\! & =\!\int_{\text{[-\textgreek{p},\textgreek{p}]}^{3}}\!\!\!\!\!\!\!\!\!d^{{\scriptscriptstyle \text{(\!3\!)}}}\mathbf{k}\frac{\varepsilon}{8\pi^{3}\left(\varepsilon^{2}\!+\!\abs{\boldsymbol{\sin k}}^{2}\right)}\label{eq:I1}\\
\mathcal{I}_{\varepsilon}^{x}\!\left(1,\!0,\!0\right)\! & =\!\int_{\text{[-\textgreek{p},\textgreek{p}]}^{3}}\!\!\!\!\!\!\!\!\!d^{{\scriptscriptstyle \text{(\!3\!)}}}\mathbf{k}\frac{\sin\!k_{x}e^{ik_{x}}}{8\pi^{3}\left(\varepsilon^{2}\!+\!\abs{\boldsymbol{\sin k}}^{2}\right)}\\
\mathcal{I}_{\varepsilon}^{0}\!\left(2,\!0,\!0\right)\! & =\!\int_{\text{[-\textgreek{p},\textgreek{p}]}^{3}}\!\!\!\!\!\!\!\!\!d^{{\scriptscriptstyle \text{(\!3\!)}}}\mathbf{k}\frac{\varepsilon e^{2ik_{x}}}{8\pi^{3}\left(\varepsilon^{2}\!+\!\abs{\boldsymbol{\sin k}}^{2}\right)},\label{eq:I3}
\end{align}
\end{subequations}

\noindent which define three complex-valued functions $f_{\varepsilon}$,
$g_{\varepsilon}$ and $h_{\varepsilon}$, respectively. In all previous
integrals $\varepsilon$ is to be taken as a complex number with a
positive infinitesimal imaginary part (guaranteeing that the LGFs
are retarded). In all three cases, analytical progress can be made
by first considering the one-dimensional integral,

\vspace{-0.3cm}

\noindent 
\begin{equation}
I_{1}\!\left(z\right)\!=\!\int_{-\pi}^{\pi}\!\!\!\!\!du\frac{1}{2\pi\left(z\!+\!\sin^{2}\!u\right)}
\end{equation}

\noindent where $z\in\mathbb{C}$. This integral can be solved by
standard contour integration in the complex variable $w\!=\!\exp\!\left(iu\right)$,
yielding

\vspace{-0.5cm}
\begin{equation}
I_{1}\!(z)\!=\!\frac{1}{\sqrt{z}\sqrt{z-1}}.
\end{equation}
Therefore,

\vspace{-0.5cm}
\begin{equation}
I_{1}^{\pm}\!\!\left(x\right)\!=\!\!\!\lim_{\eta\to0^{{\scriptscriptstyle \pm}}}\!\left[I_{1}\!\!\left(x\!+\!i\eta\right)\right]\!=\!\begin{cases}
\frac{\text{sgn}(x)}{\sqrt{x(x\!-\!1)}} & \!\!\!x\!\notin\!\left[0,1\right]\\
\mp\frac{i}{\sqrt{x(1\!-\!x)}} & \!\!\!x\!\in\!\left[0,1\right]
\end{cases}.\label{eq:I_Analytical}
\end{equation}
From Eq.\,(\ref{eq:I_Analytical}), it is easy to recognize that
all three integrals in Eqs.\,(\ref{eq:I1})-(\ref{eq:I3}) can be
written as two-dimensional $\mathbf{k}$-integrals involving $I_{1}^{{\scriptscriptstyle \pm}}\!(x)$.
Hence, they read 
\begin{figure}[t]
\begin{centering}
\hspace{-0.4cm}\includegraphics[scale=0.67]{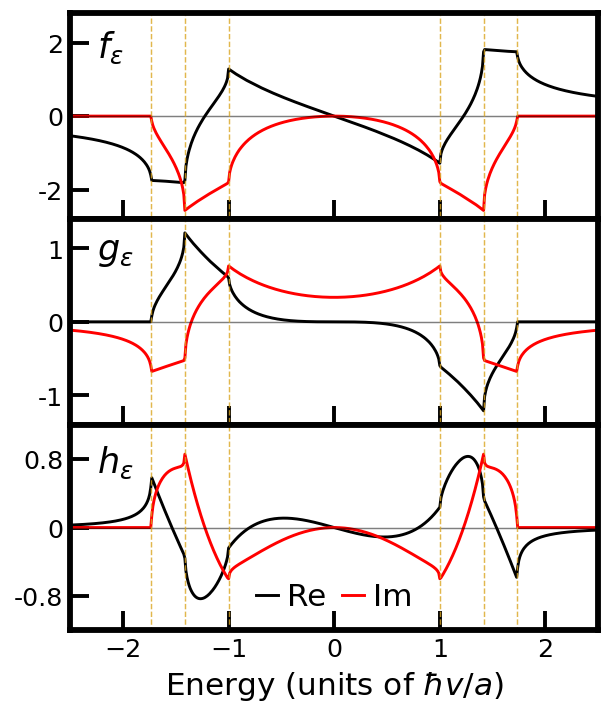}
\par\end{centering}
\vspace{-0.35cm}

\caption{\label{fig:NumericalIntegrals}Numerical integrals that define the
complex-valued functions $f_{\varepsilon}$, $g_{\varepsilon}$ and
$h_{\varepsilon}$. The node is placed at $\varepsilon\!=\!0$. Van
Hove singularities (derivative discontinuities) are marked as vertical
dimmed lines. All the curves were obtained in an regular energy mesh
of spacing $10^{{\scriptscriptstyle -5}}\hbar v/a$.}

\vspace{-0.3cm}
\end{figure}

\vspace{-0.5cm}

\begin{subequations}
\begin{align}
\!\!f_{E\pm i0^{+}}\! & =\!\frac{E}{4\pi^{2}}\int_{\text{[-\textgreek{p},\textgreek{p}]}^{2}}\!\!\!\!\!\!\!\!\!\!\!d^{{\scriptscriptstyle \text{(\!2\!)}}}\mathbf{k}\,I_{1}^{\pm}\!\left(E^{2}\!\!+\!\sin^{2}\!k_{x}\!+\!\sin^{2}\!k_{y}\right)\\
\!\!g_{E\pm i0^{+}}\! & =\!\frac{1}{4\pi^{2}}\!\!\int_{\text{[-\textgreek{p},\textgreek{p}]}^{2}}\!\!\!\!\!\!\!\!\!\!\!\!d^{{\scriptscriptstyle \text{(\!2\!)}}}\mathbf{k}\sin\!k_{x}e^{ik_{x}}\,I_{1}^{\pm}\!\left(E^{2}\!\!+\!\sin^{2}\!k_{x}\!+\!\sin^{2}\!k_{y}\right)\\
\!\!h_{E\pm i0^{+}}\! & =\!\frac{E}{4\pi^{2}}\int_{\text{[0,\textgreek{p}]}^{2}}\!\!\!\!\!\!\!\!\!\!\!d^{{\scriptscriptstyle \text{(\!2\!)}}}\mathbf{k}e^{2ik_{x}}\,I_{1}^{\pm}\!\left(E^{2}\!\!+\!\sin^{2}\!k_{x}\!+\!\sin^{2}\!k_{y}\right),
\end{align}
\end{subequations}

\vspace{-0.2cm}

\noindent where now $E$ is the (real-valued) energy parameter and
the $\pm$ denotes the sign of the energy. These two-dimensional integrals
can be numerically evaluated for an arbitrarily fine mesh of $E$.
As $\eta$ was formally taken to zero, the spectral resolution is
only limited by the spacing of this mesh. Moreover, using the symmetries
--- $f_{-E\pm i0^{+}}\!=\!-f_{E\pm i0^{+}}^{*}$, $g_{-E\pm i0^{+}}\!=\!g_{E\pm i0^{+}}^{*}$,
$h_{-E\pm i0^{+}}\!=\!-h_{E\pm i0^{+}}^{*}$ --- it is enough to
evaluate the said integrals for $E\!>\!0$. The results are shown
in Fig.\,\ref{fig:NumericalIntegrals}. 

\vspace{-0.3cm}

\section{\label{sec:Additional-Numerics}Additional Numerical Results}

\vspace{-0.1cm}

Here, we complement the KPM results shown in Sec.\ref{subsec:Average-Density-of}
for the mean DoS in the presence of a finite concentration of vacancies.
These results serve as a support for some of the claims made in the
main text.

\begin{figure}[H]
\begin{centering}
\includegraphics[scale=0.24]{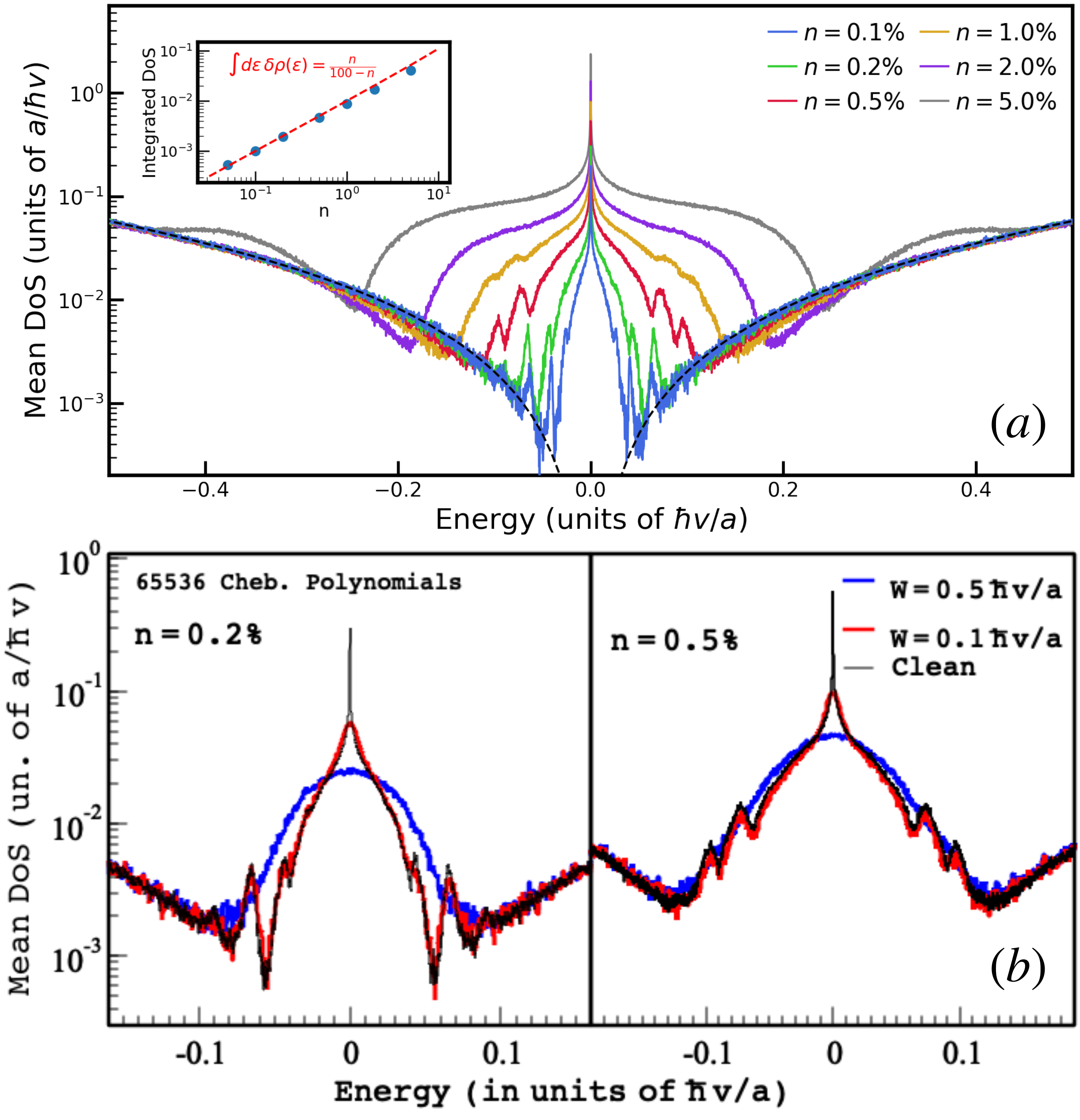}
\par\end{centering}
\vspace{-0.2cm}

\caption{\label{fig:Full-integral-of}\textbf{(a) }Full integral of the correction
do the mean DoS as a function of the vacancy concentration. Inset:
Integral of $\delta\nu(\varepsilon)\!=\!\nu(\varepsilon)\!-\!\nu_{0}(\varepsilon)$
within the region represented plotted in the main panel. \textbf{(b)}
Mean DoS calculated for two concentrations of vacancies and an additional
potential disorder.}

\vspace{-0.4cm}
\end{figure}

First, we analyze the region of enhanced mean DoS around the nodal
energy ($E\!=\!0$). For a very low vacancy concentration, the sole
feature is a sharp peak at the node, which amounts to a cumulative
contribution of all single-vacancy bound states to the intensive DoS.
As the concentration increases, one starts to see more structure,
in the form of a broadened base of this peak (as highlighted in the
main panel of Fig.\,\ref{fig:Full-integral-of}\,a). We attribute
this to a progressive hybridization of states bound to nearby vacancies
that lifts the degeneracies away from the node, thus turning these
states into long-living resonances. To corroborate this idea, we analyze
the dependence of the integrated change in the DoS within this central
region. Broadly speaking, this quantity represents the number of states
introduced near the node by the vacancies, per unit volume. In the
inset to Fig.\,\ref{fig:Full-integral-of}\,a, we show that the
integrated DoS change induced by the vacancies scales exactly as $n/(100\!-\!n)$.
This is consistent with the picture in which each missing Wannier
state introduces exactly one eigenstate in the node and one eigenstate
around the node.

Another point concerns with the robustness of the features of DoS
to microscopic details of the underlying lattice model. As we have
done for a singl\textcolor{black}{e (full) vacancy}, here also we
probe this robustness by introducing an additional Anderson potential
(of strength $W$). The KPM results for the mean DoS are shown in
Fig.\,\ref{fig:Full-integral-of}\,b. Clearly, the main features
of the DoS, \textit{i.e.}, the central enhancement and the subsidiary
peaks, remain untouched for suitably small $W$. This supports the
claim that our results will hold for a wide range of WSM systems.

$ $

$ $

\vspace{-0.6cm}

\bibliographystyle{apsrev4-2}
\bibliography{Refs}

\end{document}